\documentclass[longauth]{aa}
\usepackage{graphicx}
\usepackage{hyperref}
\usepackage{natbib}
\usepackage{float}
\usepackage{longtable}
\usepackage{csvsimple}
\usepackage{color}
\usepackage{textcomp}
\usepackage{caption}
\usepackage[retainorgcmds]{IEEEtrantools}
\usepackage[utf8]{inputenc}
\usepackage[varg]{txfonts}
\usepackage{mathabx}
\usepackage{multirow}
\usepackage{booktabs}
\usepackage{placeins}
\usepackage{amsmath}
\usepackage[caption = false]{subfig}
\usepackage{lscape}

\makeatletter
\renewcommand*{\@fnsymbol}[1]{\ensuremath{\ifcase#1\or \bigstar\or \bigstar\bigstar\or \ddagger\or
   \mathsection\or \mathparagraph\or \|\or **\or \dagger\dagger
   \or \ddagger\ddagger \else\@ctrerr\fi}}
\makeatother

\usepackage{hyperref}
\hypersetup{
    colorlinks = true,
    linkcolor = {blue}, 
    citecolor = {blue},
    urlcolor = {blue}
}

\providecommand{\bjdtdb}{\ensuremath{\rm {BJD_{TDB}}}}

\providecommand{\msun}{\ensuremath{\,M_\Sun}}
\providecommand{\rsun}{\ensuremath{\,R_\Sun}}
\providecommand{\lsun}{\ensuremath{\,L_\Sun}}

\providecommand{\me}{\ensuremath{\,M_{\rm E}}}
\providecommand{\re}{\ensuremath{\,R_{\rm E}}}
\providecommand{\fave}{\langle F \rangle}
\providecommand{\fluxcgs}{10$^9$ erg s$^{-1}$ cm$^{-2}$}

\begin{document}

    \title{Hot Rocks Survey II: 
    The thermal emission of TOI-1468 b reveals a hot bare rock
    \thanks{This article uses data from the JWST programme ID 3730.}$^{,}$\thanks{The raw and detrended photometric time-series data are available in electronic form at the CDS via anonymous ftp to cdsarc.cds.unistra.fr (130.79.128.5) or via \url{}}
    }
    
    \titlerunning{Hot Rocks Survey II: TOI-1468 b thermal emission}
    \author{
E. A. Meier Vald\'es\inst{1, 2} $^{\href{https://orcid.org/0000-0002-2160-8782}{\includegraphics[scale=0.5]{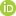}}}$, 
B.-O. Demory\inst{1,3} $^{\href{https://orcid.org/0000-0002-9355-5165}{\includegraphics[scale=0.5]{orcid.jpg}}}$,
H. Diamond-Lowe\inst{4,5} $^{\href{https://orcid.org/0000-0001-8274-6639}{\includegraphics[scale=0.5]{orcid.jpg}}}$,
J. M. Mendonça\inst{4,6,7} $^{\href{https://orcid.org/0000-0002-6907-4476}{\includegraphics[scale=0.5]{orcid.jpg}}}$,
P. C. August\inst{4} $^{\href{https://orcid.org/0000-0003-3829-8554}{\includegraphics[scale=0.5]{orcid.jpg}}}$,
M. Fortune\inst{8} $^{\href{https://orcid.org/0000-0002-8938-9715}{\includegraphics[scale=0.5]{orcid.jpg}}}$,
N. H. Allen\inst{9} $^{\href{https://orcid.org/0000-0002-0832-710X}{\includegraphics[scale=0.5]{orcid.jpg}}}$,
D. Kitzmann\inst{10} $^{\href{https://orcid.org/0000-0003-4269-3311}{\includegraphics[scale=0.5]{orcid.jpg}}}$,
A. Gressier\inst{5} $^{\href{https://orcid.org/0000-0003-0854-3002}{\includegraphics[scale=0.5]{orcid.jpg}}}$,
M. Hooton\inst{11}
$^{\href{https://orcid.org/0000-0003-0030-332X}{\includegraphics[scale=0.5]{orcid.jpg}}}$,
K. D. Jones\inst{1} $^{\href{https://orcid.org/0000-0002-2316-6850}{\includegraphics[scale=0.5]{orcid.jpg}}}$, 
L. A. Buchhave\inst{4} $^{\href{https://orcid.org/0000-0003-1605-5666}{\includegraphics[scale=0.5]{orcid.jpg}}}$,
N. Espinoza\inst{5,9} $^{\href{https://orcid.org/0000-0001-9513-1449}{\includegraphics[scale=0.5]{orcid.jpg}}}$, 
C. E. Fisher\inst{2} $^{\href{https://orcid.org/0000-0003-0652-2902}{\includegraphics[scale=0.5]{orcid.jpg}}}$,
N. P. Gibson\inst{8} $^{\href{https://orcid.org/0000-0002-9308-2353}{\includegraphics[scale=0.5]{orcid.jpg}}}$,
K. Heng\inst{12, 13, 14, 15} , 
J. Hoeijmakers\inst{16} $^{\href{https://orcid.org/0000-0001-8981-6759}{\includegraphics[scale=0.5]{orcid.jpg}}}$,
B. Prinoth\inst{16} $^{\href{https://orcid.org/0000-0001-7216-4846}{\includegraphics[scale=0.5]{orcid.jpg}}}$,
A. D. Rathcke\inst{4} $^{\href{https://orcid.org/0000-0002-4227-4953}{\includegraphics[scale=0.5]{orcid.jpg}}}$,
J. D. Eastman\inst{17} $^{\href{https://orcid.org/0000-0003-3773-5142}{\includegraphics[scale=0.5]{orcid.jpg}}}$
}

    \institute{
\label{inst:1} Center for Space and Habitability, University of Bern, Gesellschaftsstrasse 6, 3012 Bern, Switzerland \and
\label{inst:2} Department of Physics, University of Oxford, Keble Road, Oxford, OX1 3RH, UK \and
\label{inst:3} Physikalisches Institut, University of Bern, Sidlerstrasse 5, 3012
Bern, Switzerland \and
\label{inst:4} Department of Space Research and Space Technology, Technical University of Denmark, Elektrovej 328, 2800 Kgs.\,Lyngby, DK \and
\label{inst:5} Space Telescope Science Institute, 3700 San Martin Drive, Baltimore, MD 21218, USA \and
\label{inst:6} School of Physics and Astronomy, University of Southampton, Highfield, Southampton SO17 1BJ, UK \and
\label{inst:7} School of Ocean and Earth Science, University of Southampton, Southampton, SO14 3ZH, UK \and
\label{inst:8} School of Physics, Trinity College Dublin, University of Dublin, Dublin 2, Ireland \and
\label{inst:9} Department of Physics and Astronomy, Johns Hopkins University, 3400 N. Charles Street, Baltimore, MD 21218, USA \and
\label{inst:10} Space Research and Planetary Sciences, Physics Institute, University of Bern, Gesellschaftsstrasse 6, 3012 Bern, Switzerland \and
\label{inst:11} Cavendish Laboratory, JJ Thomson Avenue, Cambridge CB3 0HE, UK \and
\label{inst:12} Ludwig Maximilian University, Faculty of Physics, Scheinerstr. 1, Munich D-81679, Germany \and
\label{inst:13} ARTORG Center for Biomedical Engineering Research, University of Bern, Murtenstrasse 50, CH-3008, Bern, Switzerland \and
\label{inst:14} University College London, Department of Physics \& Astronomy, Gower St, London, WC1E 6BT, United Kingdom \and
\label{inst:15} University of Warwick, Department of Physics, Astronomy \& Astrophysics Group, Coventry CV4 7AL, United Kingdom \and
\label{inst:16} Lund Observatory, Division of Astrophysics, Department of Physics, Lund University, Box 118, 221 00 Lund, Sweden \and
\label{inst:17} Center for Astrophysics $\vert$ Harvard \& Smithsonian, 60 Garden Street, Cambridge, MA 02138, USA
}
              
\authorrunning{E.A. Meier Vald\'es et al.}
\date{Received: 14 December 2024 / Accepted: 24 March 2025}

\abstract
{Terrestrial exoplanets orbiting nearby small, cool stars known as M dwarfs are well suited for atmospheric characterisation. Given the strong XUV irradiation from M dwarf host stars, orbiting exoplanets are thought to be unable to retain primordial H/He-dominated atmospheres. However, the survivability of heavier secondary atmospheres is currently unknown.}
{The aim of the Hot Rocks Survey programme is to determine if exoplanets can retain secondary atmospheres in the presence of M dwarf hosts. Among the sample of 9 exoplanets in the programme, here we aim to determine whether TOI-1468 b has a substantial atmosphere or is consistent with a low-albedo bare rock.}
{The James Webb Space Telescope provides an opportunity to characterise the thermal emission with MIRI at 15 $\mu$m. TOI-1468 b's occultation was observed three times. We compare our observations to atmospheric models including varying amounts of CO$_{2}$ and H$_{2}$O.}
{The observed occultation depth for the individual visits are 239$\pm 52$ ppm, 341$\pm 53 $ ppm and 357$\pm 52$ ppm. A joint fit yields an occultation depth of 311$\pm 31$ ppm. The thermal emission is mostly consistent with no atmosphere and zero Bond albedo at 1.65-$\sigma$ confidence level or a blackbody at a brightness temperature of $1024 \pm 78$ K. A pure CO$_{2}$ or H$_{2}$O atmosphere with a surface pressure above 1 bar is ruled out over 3-$\sigma$.}
{Surprisingly, TOI-1468 b presents a surface marginally hotter than expected, hinting at an additional source of energy on the planet. 

It could originate from a temperature inversion, induction heating or be an instrumental artifact. The results within the Hot Rocks Survey build on the legacy of studying the atmospheres of exoplanets around M dwarfs. The outcome of this survey will prove useful to the large-scale survey on M dwarfs recently approved by the STScI.}


\maketitle

\section{Introduction}
\label{section:introduction}

In our galaxy, low-mass stars are the most abundant stars in our neighbourhood. According to the Sloan Digital Sky Survey, approximately 70\% of the stars in the Milky Way are M dwarfs \citep{Bochanski_2010}. The M spectral type is defined by a relatively small mass, spanning from the hydrogen burning limit mass of 0.08 to approximately 0.5 times the solar mass, an effective temperature between approximately 2500 to 4000 K \citep[e.g.,][]{Rajpurohit_2013}, low luminosity and strong absorption features of titanium oxide. Given their low masses, these stars burn their nuclear fuel at slower rates compared to brighter stars, thus having much longer lifetimes on the order of tens to hundreds of Gyr \citep[e.g.,][]{Tarter_2007}. In fact, all M dwarfs are still in the main sequence. The Kepler space telescope \citep{Borucki_2010} revealed that there are at least 0.5 rocky planets per M dwarf \citep{Dressing_2015}. Moreover, the occurrence rate of exoplanets with radii between 1 and 4 $R_{\Earth}$ and a period shorter than 200 days is of 2.5 planets per M dwarf \citep{Dressing_2015}. 

Interest in M dwarfs has increased in the past years due to the effort to find habitable planets and evidence of life. From an observational perspective, it is easier to find small low-mass planets orbiting small low-mass stars, since the reflex motion in the radial velocity technique and the photometric transit signal are larger given the smaller ratio in mass and radius between planet and star \citep[e.g.,][]{Shields_2016}. Terrestrial exoplanets that are suited for atmospheric characterisation orbit nearby M dwarfs (distance<30 pc). The relatively low luminosity and temperature shift the habitable zone closer to the star, increasing the frequency of short-period planets lying inside the habitable zone of its host star \citep{Nutzman_2008}. Moreover, the longevity of an M dwarf offers long timescales for planetary and biological development and evolution on orbiting planets. Projects such as TRAPPIST \citep{Gillon_2013} or SPECULOOS \citep{Jehin_2018, Delrez_2018} aim to detect terrestrial exoplanets around ultracool stars. Arguably, the most notable M-dwarf system is TRAPPIST-1 \citep{Gillon_2016}, which is host to seven Earth-sized exoplanets, with four of them lying within the habitable zone \citep{deWit_2018}. SPECULOOS, a network of robotic telescopes, recently discovered the Earth-sized exoplanet SPECULOOS-3 b \citep{Gillon_2024}, one of the most promising targets for emission spectroscopy characterisation with the James Webb Space Telescope (JWST). Other survey dedicated to the search of planets around M dwarfs include the MEarth Project \citep{Nutzman_2008}, PINES \citep{Tamburo_2022, Tamburo_2022b} and EDEN \citep{Gibbs_2020, Dietrich_2023}. The Transiting Exoplanet Survey Satellite \citep[TESS;][]{Ricker_2015}, performing a near-all-sky photometric survey has discovered many small planets around M dwarfs \citep[e.g.,][]{Crossfield_2019, Gunther_2019, Gilbert_2020, Eschen_2024}. 

While it is unlikely that rocky exoplanets around M dwarfs retain their primordial hydrogen-helium atmospheres \citep{Owen_2019}, it is plausible they could still retain secondary atmospheres as they evolve, largely sourced by outgassing of the mantle \citep{Tian_2023}. However, in general the environment around an M dwarf is not the most amenable for terrestrial exoplanet atmospheres due to strong stellar activity such as flares and intense X-ray and ultraviolet (XUV) radiation, as well as stellar winds that can erode planetary atmospheres \citep[e.g.,][]{Segura_2010, Vidotto_2013, Garraffo_2016, Garraffo_2017, Diamond_2021}. Extreme UV flux from an M dwarf can completely strip the atmosphere of a planet, leaving behind a barren rock. So far there is no conclusive detection of an atmosphere around a terrestrial exoplanet. There are hints of potential atmospheres on L 98-59 d \citep{Gressier_2024, Banerjee_2024}, L 98-59 b \citep{Bello-Arufe_2025} and 55 Cnc e \citep{Hu_2024}. However, the latter orbits an G8 V star \citep{Winn_2011}, not an M dwarf, and multiple JWST observations show flux variability, complicating a conclusive interpretation on its atmosphere \citep{Patel_2024}. Most transmission spectroscopy observations revealed flat spectra, or inconclusive results due to the degeneracy between high mean molecular weight secondary atmospheres, cloudy atmospheres or no atmospheres at all \citep[e.g.,][]{deWit_2018, Lustig_2023}. Flat transmission spectra can also result from difficulties in correcting stellar contamination from photospheric heterogeneities, such as granulation and supergranulation \citep[e.g.,][]{OSullivan_2024}. Extensive campaigns on transmission spectroscopy agree in ruling out hydrogen-helium atmospheres on rocky-type planets \citep[e.g.,][]{deWit_2018, Diamond_2018, Diamond_2020, Diamond_2022, Libby_2022, Damiano_2022, Lustig_2023, Moran_2023, Lim_2023, May_2023, Mansfield_2024, Alam_2025}.  

The outstanding issue of whether or not terrestrial worlds orbiting M dwarfs can retain secondary, higher mean molecular weight atmospheres motivated the large-size JWST proposal The Hot Rocks Survey: Testing 9 Irradiated Terrestrial Exoplanets for Atmospheres \citep[ID 3730, PI: Diamond-Lowe, Co-PI Mendonça]{Diamond_lowe_2023}. The programme aims to determine if a sample of terrestrial exoplanets orbiting M dwarfs retain secondary atmospheres or are hot bare rocks. To achieve the programme's goal, JWST measures the day-side thermal emission of the planet by observing the orbiting exoplanet as it passes behind its host star, which is called secondary eclipse or occultation. The first paper in the series revealed that the rocky planet LHS 1478 b has a shallow occultation, hinting at the possibility of an atmosphere \citep{August_2024}. 

In particular, the observations in this work are carried out by JWST with the Mid Infrared Instrument (MIRI) at 15 $\mu$m. The challenges of transmission spectroscopy for exoplanets orbiting M dwarfs, such as contamination due to unocculted star spots or faculae \citep[e.g.,][]{Pont_2008, Rackham_2017}, can be mostly mitigated by doing emission spectroscopy. By using photometry instead of spectroscopy we increase the detection signal per observation at a wavelength where the contrast between the host star's flux and thermal emission from the dayside of the planet is high. In the past, eclipse photometry has proven successful with the Spitzer Space Telescope to rule out high molecular atmospheres on the terrestrial exoplanets LHS 3844b \citep{Kreidberg_2019} and GJ 1252b \citep{Crossfield_2022}. Further motivation to observe at 15 $\mu$m is the fact that CO$_{2}$ a likely dominant species in the atmosphere for hot rocky planets \citep{Tian_2009}. Recently, JWST observations with MIRI at 15 $\mu$m probed the atmospheres of rocky exoplanets. For instance, TRAPPIST-1b's secondary eclipse was observed five times, revealing no absorption of CO$_{2}$ or other species and no heat redistribution from the dayside to the nightside of the planet, hinting at the absence of a substantial atmosphere \citep{Greene_2023, Ih_2023}. Additional observations at 12.8 $\mu$m are consistent with an airless planet or pure CO$_{2}$ resulting from temperature inversion \citep{Ducrot_2024}. The thermal emission of TRAPPIST-1c at 15 $\mu$m showed that it is most likely a bare rock with no substantial CO$_{2}$ atmosphere \citep{Zieba_2023}. MIRI Low Resolution Spectroscopy (LRS) mode has also been used to discover that GJ 1132b, Gl 486b and LTT 1445A b likely lack significant atmospheres \citep{Xue_2024, Mansfield_2024, Wachiraphan_2024}. Similarly, the emission spectrum of GJ 367b, a hot ($T_{eq}=1370 K$) sub-Earth orbiting an M dwarf, observed with MIRI LRS also ruled out a CO$_{2}$ atmosphere and no heat redistribution, being consistent with a blackbody and low albedo \citep{Zhang_2024}.    

One of the selected targets of the Hot Rocks Survey, and the subject of this paper, is TOI-1468 b, a transiting rocky exoplanet orbiting an M3.0 V star in 1.88 days. It was discovered by TESS along with its sibling transiting planet c during sectors 17, 42 and 43 \citep{Chaturvedi_2022}. The planetary nature of the signal was confirmed using radial velocity measurements from CARMENES \citep{Quirrenbach_2014} and MAROON-X \citep{Seifahrt_2018}, as well as ground-based photometric time-series. With a radius of 1.28 $R_{\Earth}$ and a mass of 3.21 $M_{\Earth}$, its resulting bulk density is consistent with a rocky composition. A striking feature about the system is that the planets are located on opposite sides of the so-called radius valley \citep{Fulton_2017}, allowing for further studies on the radius valley for small planets around M-dwarf stars. TOI-1468 c was recently observed as part of the JWST GO 3557 programme (PI: Madhusudhan) in emission with NIRISS/SOSS, NIRSpec G395H and twice with MIRI LRS to constrain atmospheric composition of mini-Neptunes around M dwarfs.   

Section \ref{section:method} presents the observing strategy with JWST MIRI, the data reduction and analysis. In Sect. \ref{section:results} we show the measured occultation depths, presenting first the joint analysis of the three visits, followed by the analysis of individual visits, accompanied by the atmospheric models. In Sect. \ref{section:discussion} we interpret the results and discuss the atmosphere of TOI-1468 b , along with possible processes producing the measured occultation depth. Finally, Section \ref{section:conclusions} provides concluding remarks and future insights into characterising terrestrial exoplanets around M dwarfs. 

\section{Methods}
\label{section:method} 

\subsection{Instrument and observing strategy}
\label{subsection:instrument and observations}

The James Webb Space Telescope (JWST) houses four main instruments: NIRSPec, NIRCam, NIRISS and MIRI. We used the Mid Infrared Instrument (MIRI) in imaging mode at 15 $\mu$m with the filter F15000W. The target star TOI-1468 was observed three times, on 29 November 2023, 1 December 2023 and 17 January 2024. Each visit lasted for 3 hours, 51 minutes and 18 seconds. Each visit consisted of 31 groups per integration and 1448 integrations per exposure, performing a single exposure time series observation in FASTR1 readout mode with an effective integration time of 9.28512 seconds. We used the SUB256 subarray to avoid saturation while having more than $\sim$20 groups per integration (as recommended by the MIRI Instrument team; S.\,Kendrew, \textit{private communication}). 

To ensure that we observe the occultation, a phase constraint was added to the observing plan. The baseline (or out-of-occultation flux) covers at least one occultation duration \citep[roughly 90 minutes,][]{Chaturvedi_2022} before and after the secondary occultation. Given the well-known ramp effect at the beginning of MIRI observations \citep[e.g.,][]{Powell_2024, Zhang_2024, Bell_2024}, we added an extra 30-minute settling time at the start of the visit. The ramp effect is identified as an excess of flux decaying---or in some cases increasing---exponentially at the beginning of an observation, presumably caused by response drift, anneal recovery by previous heating of the detector, or idle recovery between observations \citep{Dyrek_2024}. As recorded in the JWST Engineering Database under the mnemonic `IMIR\_HK\_CUR\_POS', the MIRI imaging filters used before our observations were F1800W, F2550W and F2100W, respectively. It remains to be seen if there is a persistence effect depending on the filter used before \citep[e.g.,][]{Fortune_2025}.

To avoid contamination by the transiting sibling exoplanet TOI-1468 c, the observing plan had further constraints to observe when no transit or occultation of planet c occurred. Nevertheless, we checked in our observations that there was no contaminating transit or eclipse by using a simple \texttt{batman} \citep{Kreidberg_2015} transit model of planet c. We also modelled the thermal phase-curve of planet c using Eq. (13) in \citet{Cowan_2011} in order to assess whether there is thermal contamination by the planet. We find that the maximum thermal emission of TOI-1468 c at 15 $\mu$m at an equilibrium temperature of 338 K (see Table \ref{tab:exofast}) is 56 ppm. However, at the time of our observations the phases of the planet lead to a lower contribution in thermal emission. In particular, during the first visit the lowest and highest thermal emission is 38.7 ppm and 40.4, respectively. In visit two the thermal emission varies from 53.4 to 54 ppm, while in the last visit it varies from 54.8 ppm to 55.3 ppm. The difference during the visit is of 1.7 ppm at most. Additionally, given the long orbital period of 15.54 days of planet c compared to the approximately 4 hours duration per JWST visit, its thermal phase curve can be approximated by a linear function over the course of a visit and thus is easily absorbed by the detrending model. By observing in emission rather than transmission we mitigate the possibility of contamination due to unocculted starspots. The stellar rotation period of approximately 41 days \citep{Chaturvedi_2022} is much longer than the period of the planet, so it is safe to assume the stellar flux is constant on the timescale of the observations. Additionally, optical observations have shown that TOI-1468 has weak stellar activity \citep{Chaturvedi_2022}. 

\subsection{Data reduction}
\label{subsection:reduction}

Each visit of TOI-1468 b consists of five files referred to as segments. The uncalibrated\footnote{JWST observations can be downloaded on the MAST portal at \url{https://mast.stsci.edu/portal/Mashup/Clients/Mast/Portal.html}} (.\texttt{uncal}) files were processed using the \texttt{Eureka!} reduction pipeline (version 1.1.2.dev212+gec946e65, \citeauthor{Bell_2022} \citeyear{Bell_2022}). \texttt{Eureka!} Stage 1 and 2 work as a wrapper of the official \texttt{jwst} pipeline version 1.15.1. For the data processing we used the Calibration Reference Data System (CRDS) version 11.18.4 and context 1303 reference files for the \texttt{jwst} pipeline. In Stage 1, we performed the default steps as in \texttt{jwst} to remove detector-related signatures. The steps consist of group data rescaling, creating the data quality (DQ) mask and flag bad quality pixels, flag outliers due to cosmic rays and saturation above threshold, linearity correction, dark current subtraction, jump correction, default ramp fitting and gain scale. The jump step's rejection threshold was increased to 6.0. Based on  \citet{Morrison_2023}, we drop the first and last group of each integration. Additionally, we correct for the known electromagnetic interference (EMI) noise pattern at 10 Hz and 390 Hz \citep{Bell_2024, Zhang_2024} affecting the raw MIRI data when taken with the SUB256 array. However, our dataset was not affected by the 390 Hz noise pattern, thus correcting for it induced additional noise. As such, we corrected only for the noise at 10 Hz. We performed several combinations of reduction steps including EMI correction and jump step. The reduction with EMI correction applied and skipping the jump step minimises the Median Absolute Deviation (MAD) and was selected for the next step. 

The data product of stage 1 has the suffix .\texttt{rateints}. During \texttt{Eureka!} stage 2 only a flat field correction is performed, resulting in calibrated exposures (.\texttt{calints} files). We decided to skip the \texttt{photom} step to have a better estimate of the expected photon noise. In stage 3 we crop the array into a subarray from 1 to 250 pixels in the horizontal direction and from 2 to 255 pixels in the vertical direction as seen in DS9 \citep{sao_2000} to exclude dark columns and rows at the edge of the subarray. 

We performed 5-$\sigma$ outlier rejection along the time axis for each pixel. Bad pixels are interpolated with a linear function. Pixels marked as `DO\_NOT\_USE' in the DQ array were masked. Based on visual inspection of the fits files, we select the pixel 128x128 as an initial guess for the centroid position. Then, we determined the centroid position of the star by fitting a 2D Gaussian. During the aperture photometry step we perform background subtraction. We tested different aperture sizes between 3 pixels and 20 pixels, selecting the one that minimised the MAD (see Table \ref{tab:apertures} in Appendix \ref{app: Aperture sizes}). However, the MAD of the raw light-curve might be biased by the ramp effect observed at the beginning of all the observations. Thus, we mask the first 47 minutes, corresponding to the duration of the first segment out of five, where the strongest ramp is exhibited and then compute the MAD. The selected aperture has a radius of 5 pixels for all visits. 

To estimate the background around the star, we put an annulus with an inner radius of 30 pixels and an outer radius of 50 pixels. A bigger radius is affected by the flux of nearby targets, while a smaller inner radius includes part of the so-called snowflake feature of JWST's Point Spread Function (PSF). The background is then subtracted. We did not perform $1/f$ noise correction as this type of noise only affects the NIR instruments \citep{Schlawin_2020}. We convert units from Data Number per second (DN/s) to electron count by multiplying by the gain value of 4.77 in the operational gain CRDS reference file. Stage 4 of \texttt{Eureka!} takes the calibrated files and produces the photometric light-curve. 

\subsection{Analysis}
\label{subsection:analysis}

To prepare our data, first we dealt with the ramp effect at the beginning of each observation. For this reason, we requested an additional 30 minutes of observation at the start of each visit to allow for the telescope to settle. Instead of modelling the ramp with exponential functions, we masked the first 47 minutes, corresponding to the duration of a full segment. In this fashion we avoid increasing the uncertainty in the model by having fewer parameters. Before any fitting, we normalise the flux by dividing it by its median value. Then, we removed all points deviating by more than 3$\sigma$ from the MAD. Out of 1152 data points in each visit, sigma clipping removed a total of 11, 24 and 21 from visits 1, 2 and 3, respectively. The spacecraft introduces several systematics to photometry, such as background flux level and centroid position on the detector. To analyse the calibrated light-curve, we fit a model consisting of an occultation model and systematic detrending model. 

Our detrending model is a polynomial of the form

\begin{IEEEeqnarray*}{rll}
\label{eqn:detrend}
    F_{detrend} = \ &&a_{0}+a_{1}(t-t_{0})+ a_{2}(t-t_{0})^{2}\\
                && + a_{3}(x-\overline{x})+a_{4}(y-\overline{y})+ \\
                && + a_{5}(x-\overline{x})^{2}+a_{6}(y-\overline{y})^{2}+ \\
                && + a_{7}(bg-\overline{bg})+ a_{8}(bg-\overline{bg})^{2} \IEEEyesnumber
,
\end{IEEEeqnarray*}
where $t$ is BJD\_TDB time, $t_{0}$ the start of the observation, $x$ and $y$ are centroid positions and $bg$ is the background level. The bar over a variable denotes the average value. The basis vectors a$_{i}$ are normalised. The parameter $a_{0}$ represents the baseline flux. In this work we perform a correction via polynomial regression, selecting the best combination of detrending basis vectors following \citet{Meier_2023}. The information criteria leave-one-out cross validation (see Sect. \ref{section:results} for details) favours a detrending model consisting of an intercept $a_{0}$, linear trend in time $a_{1}$ and background $a_{7}$ and ignoring centroid position. 

The occultation model based on \citet{Mandel_2002} with no limb-darkening is implemented in the \texttt{exoplanet} Python package \citep{Foreman-Mackey_2021}:

\begin{IEEEeqnarray*}{rll}
\label{eqn:occ}
    F_{occ} = f_{p}/f_{s} \cdot (occ \ model) \IEEEyesnumber
,
\end{IEEEeqnarray*}
where the parameter $f_{p}/f_{s}$ scales the occultation model and provides the occultation depth.

The measured flux is modelled as: 

\begin{IEEEeqnarray*}{rll}
\label{eqn:flux model}
    F = F_{occ}+F_{detrend} \IEEEyesnumber
.
\end{IEEEeqnarray*}

One part of the analysis consisted of a joint fit comprising the three visits. We assumed a circular orbit \citep{Chaturvedi_2022} and fixed the quadratic limb-darkening coefficients to zero. 
In our occultation model we put priors on the time of mid-transit $T_{0}$, orbital period $P$, planet-to-star radius ratio $R_{p}/R_{\star}$, impact parameter $b$, stellar radius $R_{\star}$ and mass $M_{\star}$; and fit for the occultation depth $f_{p}/f_{s}$ and coefficients of the detrending basis vectors $a_{i}$ for $i$ in \{0,1,7\} (Eq.~\ref{eqn:detrend}). The list of parameters and prior values are found in Tables 4 and 5 in \citet{Chaturvedi_2022}. The model was implemented in a Markov Chain Monte Carlo (MCMC) using the no u-turn sampler \citep[NUTS;][]{Hoffmann_2011}, an efficient variant of Hamiltonian Monte Carlo. We sampled the posterior distributions with the \texttt{PyMC3} probabilistic programming framework \citep{Salvatier_2016}, running two chains with 8000 draws and 4000 burn-in iterations with a target acceptance rate of 0.96. We checked that the chains were well mixed and ensured that the Gelman-Rubin statistic was below 1.01 for all parameters \citep{Gelman_1992}. In a second MCMC routine, we repeat the procedure, but now fitting for a different occultation depth in each visit. 

The three detrended JWST observations were then included in a set to perform a self-consistent global fit on the photometric and radial velocity observations with \texttt{EXOFASTv2} \citep{Eastman_2019} including the data presented in \citet{Chaturvedi_2022} taken by CARMENES \citep{Quirrenbach_2014}, MAROON-X \citep{Seifahrt_2018} and TESS \citep{Ricker_2015} during sector 17, 42 and 43. We included the additional TESS sector 57, observed after the publication of \citet{Chaturvedi_2022}. We fit simultaneously for both known planets in the system. The eccentricity is left free, parametrised by $\sqrt{e}\cos \omega$ and $\sqrt{e}\sin \omega$. The best-fit parameters are included in Table \ref{tab:exofast} in Appendix \ref{appendix:exofast}. The updated parameters, in particular the stellar parameters, were passed to the atmospheric models in Sect. \ref{subsection:atmospheric models} and as priors (shown in Table \ref{tab:Parameter priors}) to the MCMC routine described above, where the joint fit and individual fit were run again. 

\begin{table}[ht]
\centering\setstretch{1.0}
\caption{List of parameters in the MCMC.}
\begin{tabular}{ccc}
\hline
\hline
Parameters & Units & Priors \\
\hline 
$R_{\star}$ & [$R_{\Sun}$] & $\mathcal{N}$(0.3714, 0.01)\tablefootmark{a} \\
$M_{\star}$ & [$M_{\Sun}$] & $\mathcal{N}$(0.3527, 0.0081)\tablefootmark{a} \\
$T_{0}$ & [BJD] & $\mathcal{N}$(2458765.67755, 0.00097)\tablefootmark{a} \\
$P$ & [days] & $\mathcal{N}$(1.8805201, 0.000003)\tablefootmark{a} \\
$b$ & - & $\mathcal{N}$(0.465, 0.081)\tablefootmark{a} \\
$R_{p}/R_{\star}$ & - & $\mathcal{N}$(0.03461, 0.001)\tablefootmark{a} \\
$f_{p}/f_{s}$ & [ppm] & $\mathcal{U}$ [-100, 400] \\
$a_{i}$ & - & $\mathcal{N}$(0, 10) \\
\hline
\end{tabular}
\tablefoot{
 $\mathcal{N}$ represents a normal distribution, with the mean and standard deviation as arguments. $\mathcal{U}$ is a uniform distribution with its lower and upper bounds. $a_{i}$ are the detrending coefficients from Eq. (\ref{eqn:detrend}). \tablefoottext{a}{Priors from Table \ref{tab:exofast}.} 
}
\label{tab:Parameter priors}
\vspace{-4mm}
\end{table}

\subsection{Absolute flux calibration}
\label{subsection:flux calibration}

To perform the absolute flux calibration \citep[further details to follow in][]{Fortune_2025}, we adhere to the procedure in \citet{Gordon_2025} for the JWST Absolute Flux Calibration programme. To this purpose, we use our \texttt{.calint} produced from performing stages 1 and 2 of the standard JWST calibration pipeline, using the same settings as described in \citet{Gordon_2025}. A notable difference compared to our data reduction in Sect. \ref{subsection:reduction} is skipping the EMI correction step. The flux calibration is performed for each integration. We use \texttt{photutils} \citep{Bradley_2022} for the aperture extraction with an aperture radius of 5.69px and background annulus inner and outer radii of 8.63px and 11.45px, respectively. Those values were chosen to match the values used in the aforementioned programme \citep[see Table 4 in][]{Gordon_2025}. The conversion factor obtained in Eq.(1) in \citet{Gordon_2025} is used to convert the flux extracted from the finite aperture to the expected flux from an infinite aperture. The factor accounts for background flux due to the PSF wings. Finally, we use Eq. (3) in \citet{Gordon_2025} to convert from DN/s to millijansky (mJy).  

\subsection{Atmospheric models}
\label{subsection:atmospheric models}

In general, the measured depth of an occultation consists of a contribution by reflected light and thermal emission of the planet \citep{Mallonn_2019}:

\begin{IEEEeqnarray*}{rll}
\label{eqn:refl+therm}
    \delta_{occ} = \delta_{refl} + \delta_{therm}  \IEEEyesnumber
,
\end{IEEEeqnarray*}
where the first term is the reflected light contribution and the second is the thermal emission. The reflected light depends on the geometric albedo as follows:

\begin{IEEEeqnarray*}{rll}
\label{eqn:refl}
    \delta_{refl} = A_{g}\left(\frac{R_{p}}{a}\right)^{2}  \IEEEyesnumber
,
\end{IEEEeqnarray*}

To a first degree, the thermal emission is given by:

\begin{IEEEeqnarray*}{rll}
\label{eqn:occ depth}
    \delta_{therm} = \frac{\pi \displaystyle\int \mathcal{T}(\lambda)\mathcal{B}_{\lambda}(\lambda , T_{d}) d\lambda}{\displaystyle\int \mathcal{T}(\lambda)B_{\star}(\lambda, T_{eff}) d\lambda}  \IEEEyesnumber
,
\end{IEEEeqnarray*}
assuming isotropic radiation. $\mathcal{T}(\lambda)$ is the transmission function of the MIRI filter, $\mathcal{B}_{\lambda}(\lambda, T_{d})$ the Planck function of the planet at the dayside temperature $T_{d}$. $B_{\star}(\lambda, T_{eff})$ is the flux density from the stellar model. 

If the equilibrium temperature of the planet is lower than its effective temperature, then the planet has an internal energy source. The dayside-average brightness temperature arises from the energy balance of stellar irradiation and absorption or reradiation by the planet, given by \citep{Burrows_2014}:
\begin{IEEEeqnarray*}{rll}
\label{eqn: dayside temperature}
    T_{d} = T_{eff, \star} \sqrt{\frac{R_{\star}}{a}}(1-A_{B})^{1/4}f^{1/4} \IEEEyesnumber
,\end{IEEEeqnarray*}
where $T_{eff, \star}$ is the stellar effective temperature, $a$ the semimajor axis of the planet, $R_{\star}$ the stellar radius, $A_{B}$ the bond albedo and $f$ is the heat redistribution factor \citep[e.g.,][]{Barman_2005, Hansen_2008, Koll_2022}. In the case of zero heat redistribution from the dayside to the nightside, $f=2/3$, while for full heat redistribution $f=1/4$. For the no-atmosphere case with zero heat redistribution, we used a simplified formulation to estimate the surface temperature and outgoing flux from the planet's dayside, based on the work of \citet{Malik_2019b}.

For the atmospheric cases, we must determine the heat redistributed from the dayside to the nightside. We use the 1D atmospheric model \texttt{HELIOS} \citep{Malik_2017, Malik_2019a, Malik_2019b} to calculate the planet’s temperature structure and its synthetic emission and reflection spectra. The heat redistribution in \texttt{HELIOS} is controlled by the heat redistribution factor, which can be approximated by the scaling equation from Eq. (10) in \citet{Koll_2022}. The analytical equation depends on the longwave optical depth at the surface, the surface pressure and the equilibrium temperature.  The temperature profile of a planet's atmosphere is primarily determined by the balance between radiative, convective processes and heat redistribution. 

The model considers atmospheres with gas compositions consisting of CO$_{2}$, H$_{2}$O, or a mix of the two. It employs k-distribution tables for opacities, integrating radiative fluxes over 385 spectral bands and 20 Gaussian points. Opacities are calculated using the \texttt{HELIOS-K} \citep{Grimm_2021} code and cross-sections for CO$_{2}$ \citep{Rothman_2010} and H$_{2}$O \citep{Barber_2006}. Additionally, we incorporate Rayleigh scattering cross-sections for H$_{2}$O and CO$_{2}$ \citep{Cox_2000, Sneep_2005, Thalman_2014}. Throughout the study, we maintain a constant planetary surface albedo of 0.1 in all simulations, as this value is typical for rocky compositions \citep[e.g.,][]{Essack_2020} and a realistic minimum.

The stellar spectrum is taken from BT-Settl (CIFIST) models \citep{Allard_2011, Allard_2012} and linearly interpolated for the stellar parameters.
The model inputs found in Table \ref{tab:best fit results} and Tables 4 and 5 in \citet{Chaturvedi_2022} including the planetary parameters, semi-major axis, and stellar parameters, have associated uncertainties. In order to assess the impact of these uncertainties on the model spectra produced, we conduct 100 simulations per atmospheric scenario, where we generate random values drawn from a normal distribution for each model input parameter. 

\subsection{Independent analysis}
\label{section: independent analysis}

We used \texttt{Frida}, a JWST end-to-end pipeline, to perform an independent reduction \citep{August_2024}. \texttt{Frida} can reduce exoplanet transit and eclipse spectroscopy and photometry data products. It serves as a wrapper to stage 1 of the official JWST pipeline (\texttt{jwst}\footnote{\url{https://github.com/spacetelescope/jwst}}) to handle ramp-fitting and flat-fielding routines, and is custom-developed starting from stage 2.
\texttt{Frida} employs a time-series analysis to identify cosmic rays, flagging 5-sigma outliers in the pixel-level light curves, which are subsequently replaced using a Gaussian-smoothed interpolation over 10 integrations. The \texttt{Frida} pipeline is capable of both traditional aperture photometry and optimal spectral extraction. For the latter, we use a normalised, smoothed, median-weighted profile to define the pixel weights, effectively representing the PSF. Despite JWST's typically stable pointing, slight oscillatory drifts in the spectral position are detected with amplitudes in the thousandths of a pixel both horizontally and vertically. The smoothed PSF is accordingly adjusted for each integration to compensate for this drift.

For this dataset, we perform optimal extraction over a $20\thinspace$x$\thinspace20$ pixel grid centred on the target. We also define a "z-cut", that is a flux level below which pixels are discarded. This technique captures the intricate shape of the MIRI PSF, while excluding background contamination. The pixel grid size and z-cut level can be fine-tuned, with the cut-off defined as a fraction of the peak flux. We evaluate various configurations and determine the optimal z-cut level by examining the root mean square (RMS) and MAD of the resulting light curve. We settle on $z_{cut} = 0.02$ for the first and third observations, and $z_{cut} = 0.03$ for the second. These configurations incorporate the primary inner circle of the PSF as well as most of the secondary, petal-like patterns of the PSF.

\texttt{Frida} provides a light-curve fitting routine based on the \texttt{batman} \citep{Kreidberg_2015} code to generate occultation light curve models. We fit for the planet-to-star flux ratio $f_p/f_s$ (occultation depth) and the time of secondary eclipse $t_{secondary}$, with uniform priors covering $[0,500]\thinspace$ppm for the occultation depth and about $30\thinspace$min around the expected $t_{secondary}$, covering the range $[t_{secondary}-30\thinspace min, \ t_{secondary}+30\thinspace min]$. All other system parameters are fixed following \cite{Chaturvedi_2022}. We explore five different options to model the systematics : a first (linear) and second (quadratic) order polynomial, an exponential with a linear model, and two Gaussian Processes (GP), namely Matern 3/2 and squared exponential kernels. We then compute the Bayesian Information Criterion (BIC) \citep{Schwarz_1978} values and the reduced chi-squared statistic to compare between these methods. For the linear and quadratic models, the first $250$ integrations, corresponding to 40 minutes, are discarded to remove the full exponential ramp due to detector settling. For the exponential with a linear slope model, we remove the first $150$ integrations (24 minutes) to keep the last exponential of the ramp. Finally, the GP models are given more data to capture the shape of the noise, with only $100$ integrations (16 minutes) discarded.

Overall, we find the quadratic model performs rather poorly with a BIC difference to the best-perfoming model of 2515, 1825 and 1871 units for the first, second and third visit, respectively. Additionally, the BIC values favour the exponential with linear slope model for the first and last visit, which both exhibit a strong ramp, based on the BIC calculations. The second visit does not have such a strong detector settling ramp, and the GPs provide very flexible fits with timescales shorter than the occultation duration to fit for the correlated noise. Thus, for the second visit we chose the simpler linear model, which captures the light-curve well once the initial $250$ integrations are discarded. 

\section{Results}
\label{section:results}

First we report the findings on the individual visits in Sect. \ref{subsection:individual fit}, followed by the analysis on the joint dataset in Sect. \ref{subsection:joint fit}, compared to the expected results from the models in Sect. \ref{subsection:expected occultation depth}. The atmospheric models for different compositions of TOI-1468 b are presented in Sect. \ref{subsection:atmosphere}. The best-fit posterior parameters are shown in Table. \ref{tab:best fit results}. 

\subsection{Individual fit}
\label{subsection:individual fit}

\begin{figure*}[h!]
    \centering
    \includegraphics[width=0.95\textwidth]{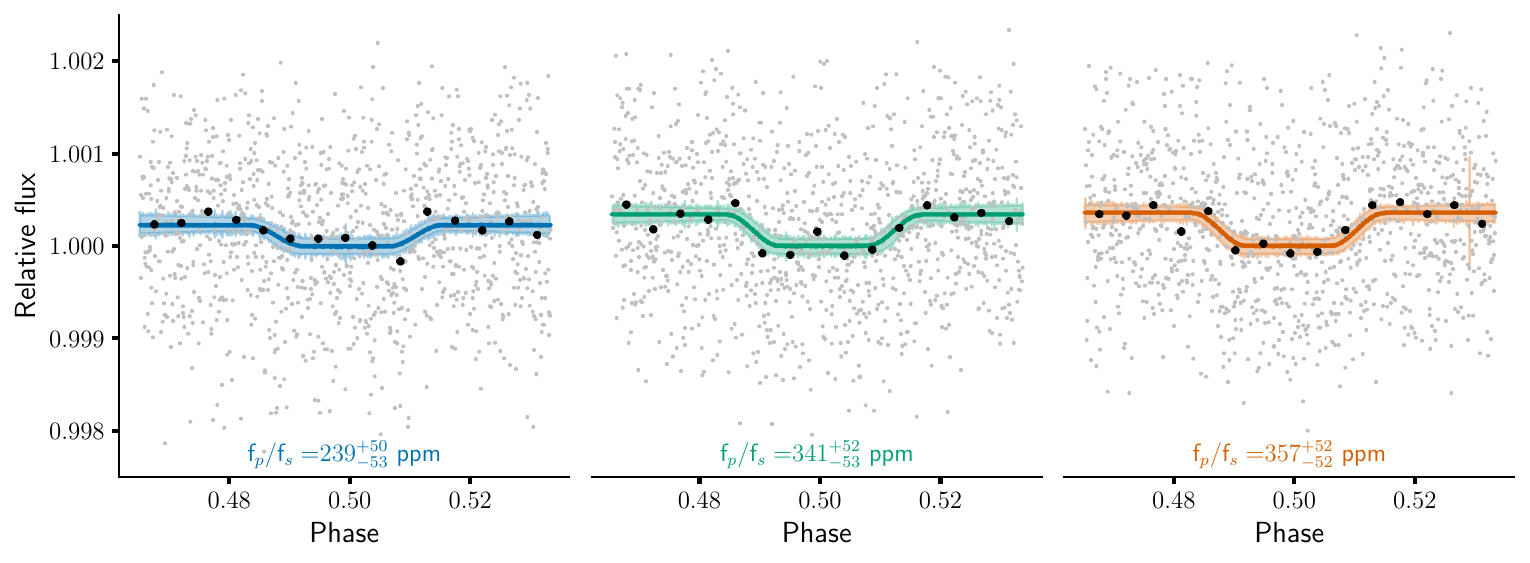}
    \caption{Light-curves of TOI-1468 b of individual fits. The left light-curve is visit 1, while the middle corresponds to visit 2 and the right one to visit 3. Gray points are detrended flux measurements. Continuous lines are the best-fit occultation models and light coloured shaded area is the 95\% high-density interval (HDI). Blue corresponds to visit 1, green to visit 2 and orange to visit 3. Black points are binned data.}
    \label{fig:individual fit}
\end{figure*}

\begin{figure}
    \centering
    \includegraphics[width=0.45\textwidth]{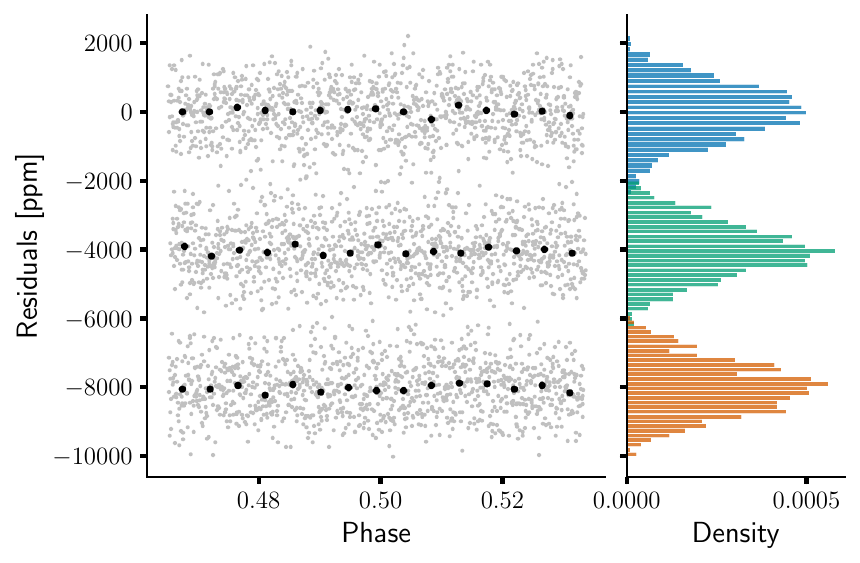}
    \caption{Flux residuals of TOI-1468 b after removing trends and occultation model. On the left panel, the top residuals are from visit 1, while the middle corresponds to visit 2 and the bottom to visit 3. Visit 2 and 3 are offset for clarity. Gray points are residual values. Black points are binned data. On the right panel the residuals density distribution is shown, aligned to its corresponding visit.}
    \label{fig:residuals individual fit}
\end{figure}

\begin{figure}
    \centering
    \includegraphics[width=0.5\textwidth]{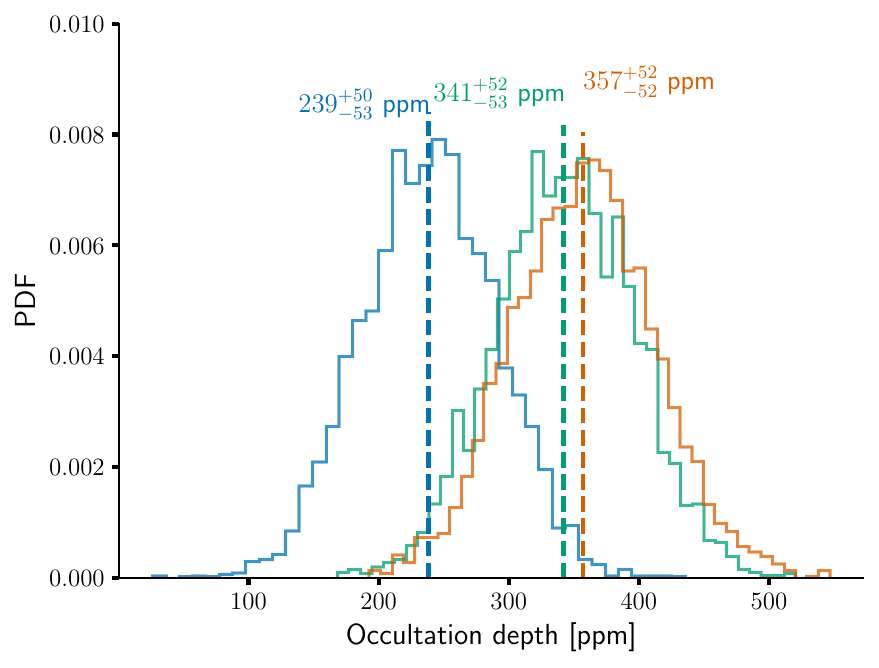}
    \caption{Posterior density distribution functions for the occultation depth parameter in visit 1 (blue), visit 2 (green) and visit 3 (orange).}
    \label{fig:pdf}
\end{figure}

For the individual observations we measure an occultation depth of 239 $^{+50}_{-53}$ppm, 341 $^{+52}_{-53}$ppm and 357 $^{+52}_{-52}$ppm for visit 1, 2 and 3, respectively. The best-fit models and detrended flux are shown in Fig. \ref{fig:individual fit}. In general, the residuals in Fig. \ref{fig:residuals individual fit} are normally distributed around zero. The unbinned residual RMS is 778, 768 and 754 ppm in visit 1, 2 and 3, respectively. The beginning of the trimmed data for visit 2 reveals an excess of flux, presumably a remnant of the ramp effect. Moreover, visit 2 exhibits a slight trend in the residuals and has the highest uncertainty in the occultation depth, although visit 1 has the highest residual RMS. Overall, the three visits agree with each other within a marginal value of 2$\sigma$ added in quadrature. In particular, visit 1 is shallower compared to the other two visits. The occultation depth in the second and third visits are consistent well within 1$\sigma$, as the posterior density distributions sampled from the MCMC depict in Fig. \ref{fig:pdf}. There is still a significant overlap in the area of the posterior distribution between all three occultations.  

For the individual fits in the independent analysis described in Sect. \ref{section: independent analysis}, we obtain $f_p/f_{s, \ 1} = 232_{-46}^{+47}\thinspace$ppm, $f_p/f_{s, \ 2} = 328_{-41}^{+37}\thinspace$ppm, and $f_p/f_{s, \ 3} = 275_{-45}^{+47}\thinspace$ppm for the first, second, and third visit respectively. Overall, the results from the independent analysis are consistent with the main analysis. Only visit 3 is marginally different at 1.4$\sigma$-significance. The reason for the difference are the detrending models implemented and the number of data points masked at the beginning of the visit.  

The average measured stellar flux described in Sect. \ref{subsection:flux calibration} during the occultation per visit has a value of 10.54 $\pm$ 0.05 mJy in every visit. We don't find evidence of variability in the stellar flux between visits within the uncertainty. In the BT\_Settl stellar model, the stellar flux weighted by the F1500W bandpass is 10.46 $\pm$ 0.62 mJy. The uncertainties propagate from the stellar distance, stellar radius, log $g$ and $T_{eff}$ in Table \ref{tab:exofast}. The values from observations and stellar model agree with each other within 0.14$\sigma$ and therefore, we do not seem necessary to adjust our stellar model by scaling it. 

\subsection{Joint fit}
\label{subsection:joint fit}

\begin{figure}
    \centering
    \includegraphics[width=0.49\textwidth]{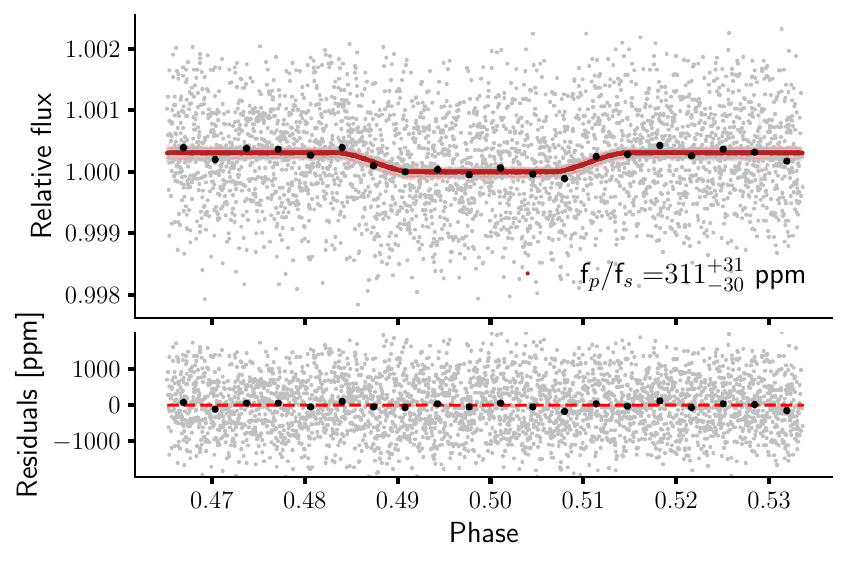}
    \caption{Phase-folded light-curves of TOI-1468 b. The top panel shows detrended flux measurements phase-folded at the planet's orbital period. Gray points are detrended flux measurements. The red line represents the best-fit joint occultation model and the light red shaded area is the 95\% highest density interval (HDI). Black dots are binned data. Bottom panel shows the residuals in ppm after removing the best-fit model, with the red dashed line displaying the zero value for visual aid.}
    \label{fig:joint fit}
\end{figure}

The joint fit reveals the detection of the occultation with a depth of 311 $^{+31}_{-30}$ppm. Table \ref{tab:best fit results} summarises the best-fit estimates of the model. Figure \ref{fig:joint fit} shows the combined visits after detrending the flux and phase-folded at the planet orbital period. The residuals do not present significant trends or signals remaining, with a residual RMS of 767 ppm. The posterior distributions and joint correlations plot do not show significant correlation between parameters (Fig. \ref{fig:corner joint} in Appendix \ref{appendix:corner plots}) except for the mean flux and slope in time, which is expected. Parallel to the observations we estimated the expected occultation depth uncertainty with the \texttt{Pandeia} engine version 4.0 \citep{Pontoppidan_2016} and RefData 3.0 using the BT\_Settl stellar model interpolated to the best-fit stellar parameters in Table \ref{tab:exofast}. Assuming that the out-of-occultation time is equal to in-occultation time, we estimated a uncertainty of 36 ppm in the joint occultation depth, while we measure an uncertainty of 31 ppm. The independent analysis described in Sect. \ref{section: independent analysis} yields a joint occultation depth of $f_p/f_{s, \ weighted} = 285\pm 25\thinspace$ppm, consistent with each other within 1$\sigma$. Moreover, the \texttt{EXOFASTv2} global fit reports an occultation depth of $298 \pm 28$ ppm, also in good agreement (see parameter $\delta _{S}$ in Table \ref{tab:exofast}). 

\begin{table}[ht]
\centering\setstretch{1.5}
\caption{Best-fit posterior estimates.}
\begin{tabular}{cc}
\hline
\hline
Parameter & Value \\
\hline 
$R_{\star}$ [$R_{\Sun}$] & $0.37229^{+0.00921}_{-0.00896}$  \\
$M_{\star}$ [$M_{\Sun}$] & $0.3530^{+0.0080}_{-0.0082}$  \\
$T_{0}$ [BJD] & $2458765.67681^{+0.00089}_{-0.00092}$  \\
$P$ [days]& $1.88051609^{+0.00000146}_{-0.00000143}$  \\
$b$ & $0.460^{+0.050}_{-0.052}$  \\
$R_{p}/R_{\star}$ & $0.0345^{+0.00099}_{-0.00099}$ \\
$f_{p}/f_{s, \ joint}$ [ppm] & $311^{+31}_{-30}$ \\
$f_{p}/f_{s, \ 1}$ [ppm] & $239^{+50}_{-53}$ \\
$f_{p}/f_{s, \ 2}$ [ppm] & $341^{+52}_{-53}$ \\
$f_{p}/f_{s, \ 3}$ [ppm] & $357^{+52}_{-52}$ \\
$log(s)$ & $-7.172^{+0.012}_{-0.012}$\\
\hline
\end{tabular}
\tablefoot{The number in the subscript in the occultation depth parameter $f_{p}/f_{s}$ refers to the corresponding visit, while $\textit{joint}$ means the orbital parameters across the three light-curves are shared, while still allowing systematics parameters to vary individually. $log(s)$ is the natural logarithm of the flux uncertainty for each measurement.
}
\label{tab:best fit results}
\vspace{-4mm}
\end{table}

Following \citet{Meier_2022}, we compare the joint fit model with a variable occultation parameter per visit and a model with zero occultation with the information criterion leave-one-out (LOO) cross-validation \citep{Vehtari_2016}. The LOO is a method for estimating the pointwise out-of-sample prediction accuracy from a Bayesian model. Compared to other information criteria such as the widely applicable information criterion (WAIC), the LOO has the advantage of being robust when the observations contain weak priors or sensitive outliers compared to other information criteria. The top-ranked model has the lowest LOO value. The higher the difference in the LOO between models, the better the top-ranked model performs. Here we follow the convention to consider a model significantly better if the $\Delta$LOO to the second-ranked model is greater than 10 \citep{McElreath_2016}. Another parameter relevant to a model comparison is the statistical weight. It can be interpreted as the estimated probability that the model will make the best predictions on future data among the considered models \citep{Yao_2018}. The values range between 0 and 1, the sum of the weights for a set of models adds up to 1. 

The models including an occultation are strongly preferred over no occultation with a combined statistical weight of 1 and $\Delta$LOO=103. The model without an occultation is ranked last with no statistical weight. Moreover, the model fitting a constant occultation parameter performs marginally better than a variable occultation with $\Delta$LOO=0.86 and a statistical weight of 0.57. 

\subsection{Expected equilibrium and brightness temperature of planet b}
\label{subsection:expected occultation depth}

To estimate the planet's equilibrium temperature, we use the stellar and orbital parameters from the \texttt{EXOFASTv2} fit in Table \ref{tab:exofast}, $T_{eff, \star} = 3376 \pm 45$ K, $R_{\star}=0.3714 \pm 0.01 \ R_{\Sun}$, $a=12.2 \pm 0.35 \ R_{\star}$, assuming zero Bond albedo and plug into Eq.(\ref{eqn: dayside temperature}). If the dayside of TOI-1468 b  does not redistribute heat to the nightside ($f=2/3$) the dayside disk-averaged temperature is $T_{d}=874 \pm 32$ K. If heat is efficiently redistributed around the sphere ($f=1/4$), the equilibrium temperature has value $T_{d}=683 \pm 32$ K. The latter is consistent with the reported value by \citet{Chaturvedi_2022} and the global \texttt{EXOFASTv2} best-fit value in Table \ref{tab:exofast}. To find the planetary brightness temperature we solve Eq.(\ref{eqn:occ depth}) numerically given the measured occultation depth of 311 $\pm$ 31 ppm including the MIRI F1500W transmission function, BT-Settl stellar model and the stellar parameters $T_{eff, \star} = 3376 \pm 45$ K and log \textit{g}=$4.846 \pm 0.025$ (Table \ref{tab:exofast}). We use the retrieval code \texttt{BeAR}, an open-source program that performs atmospheric retrievals based on \texttt{HELIOS-r2} \citep{Kitzmann_2020} with forward models in a Bayesian framework, to estimate the temperature assuming the planet as a blackbody. The best-fit for the brightness temperature of the planet yields $T_{b}=1024^{+78}_{-74}$ K. Following recent works \citep[e.g.,][]{Mansfield_2024, Xue_2024, Brandon_2024, August_2024}, we compute the ratio of measured brightness temperature to the theoretical maximum, corresponding to no heat redistribution and zero albedo scenario computed above. The ratio is 1.17 $\pm$ 0.10. To put this value into context of the Hot rocks survey, for LHS 1478 b \citet{August_2024} report a brightness temperature ratio of 0.67$\pm$0.13. In a wider context, TOI-1468 b has the highest ratio compared to other rocky exoplanets around M dwarfs observed by JWST \citep[for more details, see Table 2][]{Brandon_2024}.

\subsection{Atmosphere}
\label{subsection:atmosphere}

\begin{figure*}
    \centering
    \includegraphics[width=0.89\textwidth]{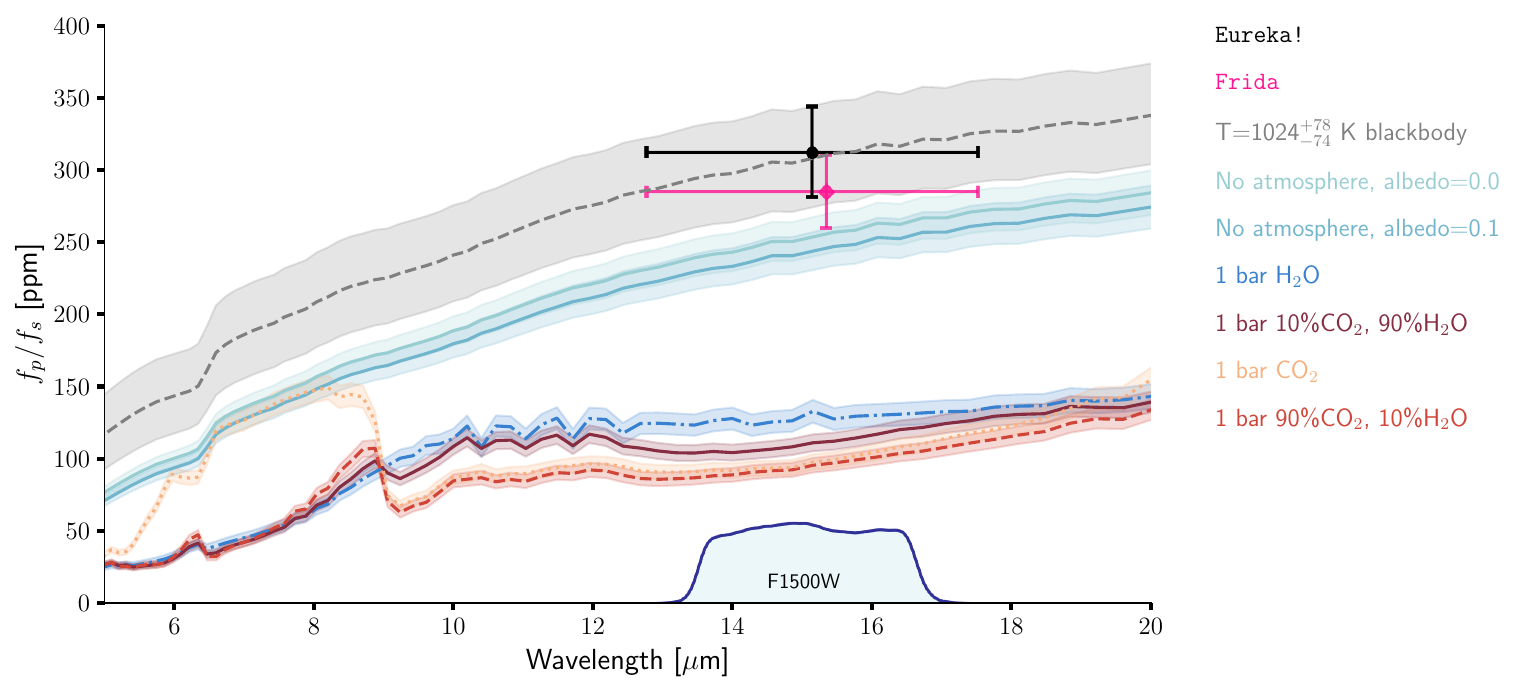}
    \caption{Emission spectra for different atmospheric scenarios. The curves show the planetary flux to stellar flux ratio in ppm as a function of wavelength in $\mu$m. In all cases we considered a  surface albedo of 0.1, heat redistribution factor based on \citet{Koll_2022} and colour-coded in the legend, ordered by the corresponding location of the model in the 15 $\mu$m bandpass. Only the no atmosphere cases have an albedo of 0 and 0.1 and zero heat redistribution. The grey dashed line and the shaded area is the blackbody emission at 1024 K and 1$\sigma$ confidence interval. The black dot is the measured joint occultation depth with its 1$\sigma$ uncertainty, while the pink diamond is the measured value from the independent analysis using \texttt{Frida}. The blue curve and light blue shaded area at the bottom is the response function of the F1500W filter.}
    \label{fig:atmospheres}
\end{figure*}

\begin{figure}
    \centering
    \includegraphics[width=0.45\textwidth]{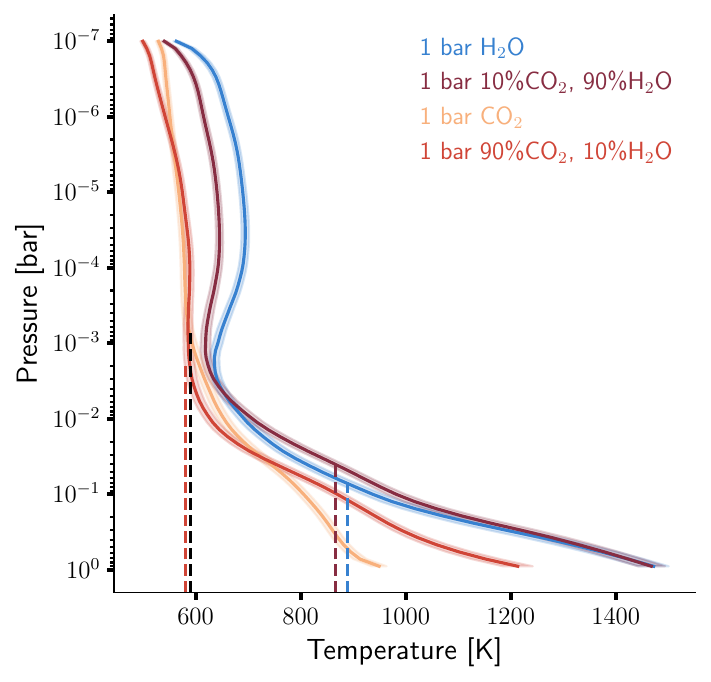}
    \caption{Temperature-pressure profiles of an atmosphere at 1 bar for varying atmospheric compositions of CO$_{2}$ and H$_{2}$O. The different models are colour-coded in the legend. The shaded region is the 1$\sigma$ confidence interval. The vertical dashed lines depict the brightness temperature of each model at 15 $\mu$m, colour-coded except for the CO$_{2}$-atmosphere for clarity.}
    \label{fig:TP}
\end{figure}

We modeled a CO$_{2}$-dominated, H$_{2}$O-dominated and a mixture of 10\% CO$_{2}$ with 90\% H$_{2}$O and 90\% CO$_{2}$ with 10\% H$_{2}$O atmospheres at a surface pressure of 1 bar. The simulations were performed for surface albedos of 0.1. Figures \ref{fig:atmospheres} and \ref{fig:TP} display the mean emission spectra and temperature profiles resulting from the simulations using \citet{Koll_2022} parametrisation of the heat redistribution factor and a surface pressure of 1 bar. In the case of no atmosphere, we present for zero and 0.1 albedo. The scenarios without atmosphere with 0 and 0.1 surface albedo are consistent with the observations at 1.65$\sigma$ and 2$\sigma$, respectively. The cases with a CO$_{2}$- or H$_{2}$O-dominated atmosphere or a mix between the two species at 1 bar or above are ruled out over 3$\sigma$. 

Among the considered models, the observations are mostly consistent with a blackbody emitter at a median temperature of 1024 K or an bare rock devoid of atmosphere. To increase the energy budget on the dayside and produce higher emission, we explore the scenario of no heat redistribution. Heat transport on tidally-locked exoplanets depends on multiple mechanisms, but strongly on the presence of an atmosphere \citep[e.g.,][]{Wordsworth_2015}. For instance, even a thin atmosphere can redistribute heat to some extent to the nightside \citep{Pierrehumbert_2019}. Solid mantles, on the other side, redistribute heat inefficiently \citep[e.g.,][]{Tobias_2021}. Hints of inefficient heat redistribution on a tidally-locked exoplanet have been observed on LHS 3844 b \citep{Kreidberg_2019}.
In this work, at 15 $\mu$m all models for no redistribution have an approximately 50 ppm higher emission than for full redistribution, but are still slightly shallower compared to the observations (see Fig. \ref{fig:extreme f cases}). 

The features in an emission spectra are linked to changes in the temperature profile, shown in Fig. \ref{fig:TP}.  The atmosphere with lower CO$_{2}$ levels are more transparent than the CO$_{2}$-dominated atmosphere. A large contribution to the outgoing emission at 15 $\mu$m comes from the deeper layers of the atmosphere.

\section{Discussion}
\label{section:discussion}

The measured thermal emission is higher than expected from theoretical models. Moreover, the observed occultation is inconsistent with a cloud-free CO$_{2}$ or H$_{2}$O atmosphere at surface pressures above 1 bar at 15$\mu$m. Given the effective temperature is above the dayside temperature assuming zero heat redistribution suggests that TOI-1468 b is a hot and airless bare rock, consistent at 1.65 sigma or a blackbody at an effective temperature of around 1024 K. The fact that $T_{b}$ is slightly higher than $T_{d}$ might hint to an additional source of energy on the planet. Here we discuss possible mechanisms that could be causing the deep occultation observed. 

\subsection{Thermal inversion}
\label{subsection: thermal inversion}

In case TOI-1468 b has an atmosphere, it may contain a very efficient absorber of stellar radiation, leading to a temperature inversion. In general, when the temperature profile gets hotter as pressure increases we get an absorption feature, but if there is a thermal inversion in the atmosphere, we get an emission feature. If the atmosphere contains a species that absorbs efficiently the stellar radiation, that might cause an inversion, leading to a higher thermal emission. For instance, the ozone layer on Earth contributes to a thermal inversion in the atmosphere. Silicate atmospheres contain different species that are strong absorbers at IR, notably Na and SiO in lava planets \citep{Zilinskas_2022}. Models show that for planets with a mass between 1 and 10 $M_{\Earth}$ and surface temperature below 2096 K the partial pressure is dominated by Na, O$_{2}$ and O, while SiO dominates at higher surface temperature \citep{Miguel_2011}. The estimated temperature of 1000 K on TOI-1468 b discards SiO as possible absorber, but Na is a possible candidate to dominate the atmosphere and produce an inversion layer. Certain atmospheric conditions might lead to the formation of Na clouds. In particular, the Na condensates NaCl and Na$_{2}$S are stable at surface pressures of 1 bar and above \citep[see Fig. 3 in][]{Marley_2013}. If the clouds are thick enough, they might generate variation in the surface albedo and imprint a signal in the dayside thermal emission \citep{Castan_2011}, potentially explaining the observed thermal emission on TOI-1468 b.  

\subsection{Induction heating}

Stellar mass models \citep{Shields_2016} place the star between the partially to fully convective regime ($M_{\Sun}<0.35$). It has been observed that several fully convective M dwarfs have strong magnetic fields on the order of 1000 G \citep{Vidotto_2014}. Under special conditions, the varying magnetic field on the orbiting planet generates induction heating, increasing the energy budget. Induction heating can increase outgassing and even form subsurface magma oceans on rocky exoplanets. A necessary condition for induction heating to occur is that the normal of the orbital plane is inclined with respect to the stellar rotation axis, or that the magnetic field is misaligned to the stellar rotation axis \citep{Kislyakova_2018}. It is possible that a strong variable magnetic field on TOI-1468 is heating up planet b. The effect can be quantified with spectrophotometers such as SPIRou \citep{Donati_2020}. Zeeman-Doppler Imaging (ZDI) maps the magnetic field and its geometry, providing evidence on induction heating. The challenge is that the rotation period of the star is around 40 days and SPIRou cannot perform continuous observations for such a long time frame. An alternative is to model the magnetohydrodynamics of the system, which is out of the scope of this paper. 

\subsection{Instrumental artifact}

The origin could be at the JWST data reduction. The MIRI instrument has particular added challenges during data reduction, such as the EMI-noise. Various data-processing algorithms, such as \texttt{Eureka!} \citep{Bell_2022}, are being employed for processing JWST observations from raw data to a spectrum. The diverse approaches show that there is not an optimal way to process JWST data yet. Fortunately, the scientific community has been quick at studying the instruments performance, reporting issues and how to overcome them \citep[e.g.][]{Dyrek_2024, Libralato_2024, Morrison_2023}. Different reduction procedures can lead to vastly different interpretations of the observations. It is plausible that the data reduction is responsible for the deep occultation measured. To ensure that this is not the case, we performed three independent data reduction and analysis (only one is presented in Sect. \ref{section: independent analysis}), being careful to not use the same pipeline or data processing tools. The derived occultation depth agreed with each other within 2$\sigma$, with no analysis reporting an occultation depth below 200 ppm. Consequently, it is unlikely that the data reduction presented in Sect. \ref{subsection:reduction} is inducing the excessively deep occultation.

Another possibility is that a specific unaccounted source of noise in MIRI is affecting the observations. Visual inspection of the individual frames shows a vertical gradient on the electron counts. Specifically, the counts in the lower region of the subarray are lower, gradually increasing in the vertical direction. This might not impact our analysis as the selected background region for subtraction encompasses a region far from the top and bottom. We find marginal evidence of correlated noise in the residuals (see Appendix \ref{app: rms}). Nevertheless, if multiple MIRI observations yield a deeper occultation depth systematically, then it might hint at a currently unaccounted systematic associated with the instrument.

\subsection{Tidal heating}
\label{subsection: tidal heating}

A possible explanation is that TOI-1468 b  moves in an eccentric orbit. Notably, TOI-1468 b 's orbit could be perturbed by the presence of its transiting sibling planet TOI-1468 c. An eccentric orbit produces internal heat due to the tidal effects on the planet caused by the star or other planets \citep{Correia_Laskar_2010}. Substantial internal energy of the planet can result as orbital energy is converted into tidal energy during the planet's evolution \citep{Perryman_2018}. A planet in an eccentric orbit will experience variable tidal forces as it moves closer and farther from its host star, being periodically deformed. If the planet experiences a rapid circularisation, the dissipated energy can exceed the planet's internal energy, making the planet glow \citep{Perryman_2018}, translating to a higher thermal emission. For instance, GJ 876d \citep{Rivera_2005}, the first sub-Neptune discovered and orbiting an M4.0V star, seems to have experienced substantial tidal heating, whose budget exceeds the required power to melt a silicate mantle \citep{Valencia_2007}.    
To challenge this hypothesis, we included the eccentricity as free parameter in the joint fit with  \texttt{EXOFASTv2} \citep{Eastman_2019}. However, the best-fit parameters reveal that the  eccentricity for TOI-1468 b is 0.0099$^{+0.018}_{-0.005}$ (see Table \ref{tab:exofast}) and thus consistent with zero, inferring that tidal heating is an unlikely explanation.  

\subsection{System contamination}

The thermal emission from atmospheric models depends strongly on the stellar model. For instance, it is well known that M dwarfs are active stars \citep[e.g.,][]{Diamond_2021}, as flares and intense XUV activity are common. However, TOI-1468 seems to have weak activity at optical wavelengths \citep{Chaturvedi_2022}, but activity at high energies is still possible \citep[e.g.,][]{Loyd_2018}. The issue is that the host star has been poorly studied in the past. In fact, most of the stellar and orbital parameters originate from the discovery paper \citep{Chaturvedi_2022}. Follow-up observations of the host star might provide insightful information to the models. Different stellar parameters will affect the theoretical emission spectra and atmospheric retrieval, potentially changing the fitting atmospheric composition to the observations of TOI-1468 b. 
While stellar activity might not significantly affect the temperature profile, it could influence the chemistry in the atmosphere.  

It is also possible that the sibling transiting planet c is contaminating the TOI-1468 b light-curve with its thermal emission. To test this, we included the thermal phase-curve of TOI-1468 c using the best-fit parameters in Table \ref{tab:exofast} in our MCMC routines. During the observations and given the long orbital period of planet c compared to the observation timescale, the thermal phase-curve can be approximated by a linear function and does not have an impact in the resulting occultation depth of TOI-1468 b. Ultimately, we find no evidence of thermal contamination by planet c.

\subsection{Stellar spectrum models}
\label{subsection: stellar spectrum models}

Another factor related to the host star that could affect the modelled emission spectra is the stellar model. A discrepancy between the measured stellar flux and the expected flux from the stellar model indicates that an adjustment to the model is required \citep[see e.g.,][]{Fauchez_2025}. If the measured flux is higher than the flux from the model, then scaling the stellar spectrum up translates into a lower modelled thermal emission. An excess of flux from the stellar model compared to the measurements requires to scale the spectrum down, producing a higher modelled thermal emission. In our case, the measured flux and expected stellar flux using the interpolated BT\_Settl model agree with each other within 0.14$\sigma$. In particular, the measured flux is 1.2$\pm$ 6.2\% marginally higher compared to the BT\_Settl model. This is similar to the results for TRAPPIST-1 c \citep{Zieba_2023}, where both fluxes are consistent within 1\%. In contrast, \citet{Greene_2023} found a 13\% excess in the measured flux of TRAPPIST-1 compared to the PHOENIX model. For our work, no rescaling is necessary and we conclude that our BT\_Settl stellar model is not causing the atmospheric models to yield a lower thermal emission compared to the observations. However, it has to be noted that our comparison applies only to the 15 $\mu$m bandpass. The stellar model can still be incorrect or require adjustment at other wavelengths and thus be the source of discrepancy compared to observations.  

\section{Conclusions}
\label{section:conclusions}

The abundance of M dwarfs in our galaxy and the feasibility of finding and characterizing a rocky exoplanet around such a star ignited interest in M dwarf systems \citep{Bochanski_2010, Gould_2003, Dressing_2015}. So far, there is no conclusive detection of an atmosphere around a terrestrial exoplanet orbiting an M dwarf. There are, however, hints of an atmosphere, such as the tentative sulphur-rich atmosphere around L 98-59 d \citep{Banerjee_2024, Gressier_2024}. Motivated by this issue, the Hot Rocks Survey aims to determine whether terrestrial exoplanets around M dwarfs have a thick, CO$_{2}$-rich atmosphere or are bare rocks. Here we present the three observations of the occultation of TOI-1468 b  \citep{Chaturvedi_2022}, a rocky exoplanet (1.4$R_{\Earth}$) orbiting an M3V dwarf. 

The analysis yields a joint occultation depth of 311 $\pm$ 31 ppm. From the individual visits we find an occultation depth of 239 $\pm$ 52 ppm, 341 $\pm$ 53 ppm and 357 $\pm$ 52 ppm. The derived dayside brightness temperature of 1024$^{+78}_{-74}$ K is slightly higher than the dayside temperature with no atmosphere and zero surface albedo at 1.65$\sigma$ and thus statistically not significant. 
The atmospheric models show that if the planet has an atmosphere, it does not have a significant amount of CO$_{2}$ or H$_{2}$O. It is also plausible that TOI-1468 b is a hot bare rock with no atmosphere, but this is not conclusive, as the process causing the deeper than predicted occultation depth remains unsolved.

The origin of the high dayside temperature is currently unknown, but we propose both instrumental and astrophysical hypotheses. In case the signal is astrophysical in nature, this would be an intriguing result, suggesting there is an unaccounted source of energy heating the planet. In the case of harbouring an atmosphere, it may possess an efficient absorber of stellar radiation, leading to a temperature inversion in the atmosphere and imprinting a stronger signal in thermal emission. Despite the presence of the sibling transiting planet c that could perturb the system and contribute to tidal heating on planet b, the joint fit gives a near circular orbit, ruling out tidal heating as a possibility. The fact that M dwarfs have been observed to have strong magnetic fields presents the possibility of induction heating. Depending on the system's architecture, a variable magnetic field strength due to the motion of the exoplanet can provide enough energy to even melt the surface of a rocky exoplanet \citep{Kislyakova_2018}. The contribution of induction heating can be a strong source of additional energy in the exoplanet \citep{Driscoll_2015}.

The results presented here motivate for future observations at different wavelength channels, as the MIRI imaging at 15 $\mu$m provides a single photometric point. 
Alternatively, by measuring a full phase curve of TOI-1468 b we can obtain precise observations of the day- and nightside emission, which lead unambiguously to the detection or to the lack of an atmosphere \citep{Hammond_2024}. The work presented here showcases the potential of the Hot Rocks Survey to deepen our understanding of atmospheres on terrestrial exoplanets around M dwarfs. The interest in such objects is reflected by the initiative to search for atmospheres of rocky M-dwarf exoplanets recently approved by the STScI \citep{Redfield_2024}. The large-scale survey will use approximately 500 hours of Director's Discretionary Time on the JWST to observe around a dozen exoplanets in nearby systems, while roughly 250 HST orbits will complement the survey with ultraviolet observations to characterise the activity of the host stars. The results provided by the Hot Rocks Survey will serve as a valuable pathway to this large-scale survey. 

\begin{acknowledgements}
We are grateful to the anonymous referee for the careful
reading and thoughtful suggestions that improved this paper.
Based on observations with the NASA/ESA/CSA James Webb Space Telescope obtained at the Space Telescope Science Institute, which is operated by the Association of Universities for Research in Astronomy, Incorporated, under NASA contract NAS5-03127.
EMV acknowledges support from the Centre for Space and Habitability (CSH). This work has been carried out within the framework of the National Centre of Competence in Research PlanetS supported by the Swiss National Science Foundation under grant 51NF40\_205606. EMV acknowledges financial support from the Swiss National Science Foundation (SNSF) Postdoc.Mobility Fellowship under grant no. P500PT\_225456/1.
B.-O. D. acknowledges support from the Swiss State Secretariat for Education, Research and Innovation (SERI) under contract number MB22.00046.
HDL and PCA acknowledges support from the Carlsberg Foundation, grant CF22-1254.
JMM acknowledges support from the Horizon Europe Guarantee Fund, grant EP/Z00330X/1.
NHA acknowledges support by the National Science Foundation Graduate Research Fellowship under Grant No. DGE1746891.
CEF acknowledges support from the European Research Council (ERC) under the European Union's Horizon 2020 research and innovation program under grant agreement no. 805445.
NPG gratefully acknowledges support from Science Foundation Ireland and the Royal Society through a University Research Fellowship (URF\textbackslash R\textbackslash 201032).
HJH acknowledges support from eSSENCE (grant number eSSENCE@LU 9:3), The Crafoord foundation and the Royal Physiographic Society of Lund, through The Fund of the Walter Gyllenberg Foundation.
BP acknowledges support from the Walter Gyllenberg Foundation. 
ADR acknowledges support from the Carlsberg Foundation, grant CF22-1548.
We acknowledge the input from the following individuals to the GO 3730 proposal: Can Akin, Andrea Guzman Mesa, Nicholas Borsato, Adam Burgasser, Meng Tian, Mette Baungaard.
We thank Dr. Yamila Miguel for insightful discussions on atmospheric thermal inversion.
We thank Dr.\,Jen Winters for important conversations on the potential multiplicity of the host star.
The authors thank Suzanne Aigrain for discussions on stellar activity and data analysis.
This research made use of \texttt{exoplanet} \citep{Foreman-Mackey_2021} and its
dependencies \citep{exoplanet:agol20, exoplanet:arviz, exoplanet:astropy13,
exoplanet:astropy18, exoplanet:kipping13, exoplanet:luger18, Salvatier_2016,
exoplanet:theano}. We acknowledge the use of further software: \texttt{batman} \citep{Kreidberg_2015}, \texttt{NumPy} \citep{Harris_2020}, \texttt{matplotlib} \citep{Hunter_2007}, \texttt{corner} \citep{corner_2016}, \texttt{astroquery} \citep{Ginsburg_2019}  and \texttt{scipy} \citep{scipy_2020}. This work made use of \texttt{seaborn} \citep{Waskom_2021} to generate the colourblind palette. 

\end{acknowledgements}

\bibliographystyle{aa}
\setcitestyle{authoryear,open={(},close={)}}
\bibliography{reference}

\begin{thebibliography}{144}
\expandafter\ifx\csname natexlab\endcsname\relax\def\natexlab#1{#1}\fi

\bibitem[{{Agol} {et~al.}(2020){Agol}, {Luger}, \& {Foreman-Mackey}}]{exoplanet:agol20}
{Agol}, E., {Luger}, R., \& {Foreman-Mackey}, D. 2020, \aj, 159, 123

\bibitem[{Alam {et~al.}(2024)Alam, Gao, Adams~Redai, Wallack, Wogan, Aguichine, Dattilo, Alderson, Batalha, Batalha, Kirk, L{\'o}pez-Morales, Meech, Moran, Teske, Wakeford, \& Wolfgang}]{Alam_2025}
Alam, M.~K., Gao, P., Adams~Redai, J., {et~al.} 2024, \aj, 169, 15

\bibitem[{Allan(1966)}]{Allan_1966}
Allan, D. 1966, Proceedings of the IEEE, 54, 221

\bibitem[{{Allard} {et~al.}(2011){Allard}, {Homeier}, \& {Freytag}}]{Allard_2011}
{Allard}, F., {Homeier}, D., \& {Freytag}, B. 2011, in Astronomical Society of the Pacific Conference Series, Vol. 448, 16th Cambridge Workshop on Cool Stars, Stellar Systems, and the Sun, ed. C.~{Johns-Krull}, M.~K. {Browning}, \& A.~A. {West}, 91

\bibitem[{{Allard} {et~al.}(2012){Allard}, {Homeier}, \& {Freytag}}]{Allard_2012}
{Allard}, F., {Homeier}, D., \& {Freytag}, B. 2012, Philosophical Transactions of the Royal Society of London Series A, 370, 2765

\bibitem[{{Astropy Collaboration} {et~al.}(2018){Astropy Collaboration}, {Price-Whelan}, {Sip{\H o}cz}, {G{\"u}nther}, {Lim}, {Crawford}, {Conseil}, {Shupe}, {Craig}, {Dencheva}, {Ginsburg}, {VanderPlas}, {Bradley}, {P{\'e}rez-Su{\'a}rez}, {de Val-Borro}, {Aldcroft}, {Cruz}, {Robitaille}, {Tollerud}, {Ardelean}, {Babej}, {Bach}, {Bachetti}, {Bakanov}, {Bamford}, {Barentsen}, {Barmby}, {Baumbach}, {Berry}, {Biscani}, {Boquien}, {Bostroem}, {Bouma}, {Brammer}, {Bray}, {Breytenbach}, {Buddelmeijer}, {Burke}, {Calderone}, {Cano Rodr{\'{\i}}guez}, {Cara}, {Cardoso}, {Cheedella}, {Copin}, {Corrales}, {Crichton}, {D'Avella}, {Deil}, {Depagne}, {Dietrich}, {Donath}, {Droettboom}, {Earl}, {Erben}, {Fabbro}, {Ferreira}, {Finethy}, {Fox}, {Garrison}, {Gibbons}, {Goldstein}, {Gommers}, {Greco}, {Greenfield}, {Groener}, {Grollier}, {Hagen}, {Hirst}, {Homeier}, {Horton}, {Hosseinzadeh}, {Hu}, {Hunkeler}, {Ivezi{\'c}}, {Jain}, {Jenness}, {Kanarek}, {Kendrew}, {Kern}, {Kerzendorf}, {Khvalko}, {King}, {Kirkby}, {Kulkarni},
  {Kumar}, {Lee}, {Lenz}, {Littlefair}, {Ma}, {Macleod}, {Mastropietro}, {McCully}, {Montagnac}, {Morris}, {Mueller}, {Mumford}, {Muna}, {Murphy}, {Nelson}, {Nguyen}, {Ninan}, {N{\"o}the}, {Ogaz}, {Oh}, {Parejko}, {Parley}, {Pascual}, {Patil}, {Patil}, {Plunkett}, {Prochaska}, {Rastogi}, {Reddy Janga}, {Sabater}, {Sakurikar}, {Seifert}, {Sherbert}, {Sherwood-Taylor}, {Shih}, {Sick}, {Silbiger}, {Singanamalla}, {Singer}, {Sladen}, {Sooley}, {Sornarajah}, {Streicher}, {Teuben}, {Thomas}, {Tremblay}, {Turner}, {Terr{\'o}n}, {van Kerkwijk}, {de la Vega}, {Watkins}, {Weaver}, {Whitmore}, {Woillez}, {Zabalza}, \& {Astropy Contributors}}]{exoplanet:astropy18}
{Astropy Collaboration}, {Price-Whelan}, A.~M., {Sip{\H o}cz}, B.~M., {et~al.} 2018, \aj, 156, 123

\bibitem[{{Astropy Collaboration} {et~al.}(2013){Astropy Collaboration}, {Robitaille}, {Tollerud}, {Greenfield}, {Droettboom}, {Bray}, {Aldcroft}, {Davis}, {Ginsburg}, {Price-Whelan}, {Kerzendorf}, {Conley}, {Crighton}, {Barbary}, {Muna}, {Ferguson}, {Grollier}, {Parikh}, {Nair}, {Unther}, {Deil}, {Woillez}, {Conseil}, {Kramer}, {Turner}, {Singer}, {Fox}, {Weaver}, {Zabalza}, {Edwards}, {Azalee Bostroem}, {Burke}, {Casey}, {Crawford}, {Dencheva}, {Ely}, {Jenness}, {Labrie}, {Lim}, {Pierfederici}, {Pontzen}, {Ptak}, {Refsdal}, {Servillat}, \& {Streicher}}]{exoplanet:astropy13}
{Astropy Collaboration}, {Robitaille}, T.~P., {Tollerud}, E.~J., {et~al.} 2013, \aap, 558, A33

\bibitem[{{August} {et~al.}(2025){August}, {Buchhave}, {Diamond-Lowe}, {Mendon{\c{c}}a}, {Gressier}, {Rathcke}, {Allen}, {Fortune}, {Jones}, {Meier Vald{\'e}s}, {Demory}, {Espinoza}, {Fisher}, {Gibson}, {Heng}, {Hoeijmakers}, {Hooton}, {Kitzmann}, {Prinoth}, {Eastman}, \& {Barnes}}]{August_2024}
{August}, P.~C., {Buchhave}, L.~A., {Diamond-Lowe}, H., {et~al.} 2025, \aap, 695, A171

\bibitem[{{Banerjee} {et~al.}(2024){Banerjee}, {Barstow}, {Gressier}, {Espinoza}, {Sing}, {Allen}, {Birkmann}, {Challener}, {Crouzet}, {Haswell}, {Lewis}, {Lewis}, \& {Yang}}]{Banerjee_2024}
{Banerjee}, A., {Barstow}, J.~K., {Gressier}, A., {et~al.} 2024, \apjl, 975, L11

\bibitem[{{Barber} {et~al.}(2006){Barber}, {Tennyson}, {Harris}, \& {Tolchenov}}]{Barber_2006}
{Barber}, R.~J., {Tennyson}, J., {Harris}, G.~J., \& {Tolchenov}, R.~N. 2006, \mnras, 368, 1087

\bibitem[{Barman {et~al.}(2005)Barman, Hauschildt, \& Allard}]{Barman_2005}
Barman, T.~S., Hauschildt, P.~H., \& Allard, F. 2005, \apj, 632, 1132

\bibitem[{{Bell} {et~al.}(2022){Bell}, {Ahrer}, {Brande}, {Carter}, {Feinstein}, {Caloca}, {Mansfield}, {Zieba}, {Piaulet}, {Benneke}, {Filippazzo}, {May}, {Roy}, {Kreidberg}, \& {Stevenson}}]{Bell_2022}
{Bell}, T., {Ahrer}, E.-M., {Brande}, J., {et~al.} 2022, JOSS, 7, 4503

\bibitem[{{Bell} {et~al.}(2024){Bell}, {Crouzet}, {Cubillos}, {Kreidberg}, {Piette}, {Roman}, {Barstow}, {Blecic}, {Carone}, {Coulombe}, {Ducrot}, {Hammond}, {Mendon{\c{c}}a}, {Moses}, {Parmentier}, {Stevenson}, {Teinturier}, {Zhang}, {Batalha}, {Bean}, {Benneke}, {Charnay}, {Chubb}, {Demory}, {Gao}, {Lee}, {L{\'o}pez-Morales}, {Morello}, {Rauscher}, {Sing}, {Tan}, {Venot}, {Wakeford}, {Aggarwal}, {Ahrer}, {Alam}, {Baeyens}, {Barrado}, {Caceres}, {Carter}, {Casewell}, {Challener}, {Crossfield}, {Decin}, {D{\'e}sert}, {Dobbs-Dixon}, {Dyrek}, {Espinoza}, {Feinstein}, {Gibson}, {Harrington}, {Helling}, {Hu}, {Iro}, {Kempton}, {Kendrew}, {Komacek}, {Krick}, {Lagage}, {Leconte}, {Lendl}, {Lewis}, {Lothringer}, {Malsky}, {Mancini}, {Mansfield}, {Mayne}, {Evans-Soma}, {Molaverdikhani}, {Nikolov}, {Nixon}, {Palle}, {Petit dit de la Roche}, {Piaulet}, {Powell}, {Rackham}, {Schneider}, {Steinrueck}, {Taylor}, {Welbanks}, {Yurchenko}, {Zhang}, \& {Zieba}}]{Bell_2024}
{Bell}, T.~J., {Crouzet}, N., {Cubillos}, P.~E., {et~al.} 2024, Nature Astronomy, 8, 879

\bibitem[{{Bello-Arufe} {et~al.}(2025){Bello-Arufe}, {Damiano}, {Bennett}, {Hu}, {Welbanks}, {MacDonald}, {Seligman}, {Sing}, {Tokadjian}, {Oza}, \& {Yang}}]{Bello-Arufe_2025}
{Bello-Arufe}, A., {Damiano}, M., {Bennett}, K.~A., {et~al.} 2025, \apjl, 980, L26

\bibitem[{{Bochanski} {et~al.}(2010){Bochanski}, {Hawley}, {Covey}, {West}, {Reid}, {Golimowski}, \& {Ivezi{\'c}}}]{Bochanski_2010}
{Bochanski}, J.~J., {Hawley}, S.~L., {Covey}, K.~R., {et~al.} 2010, \aj, 139, 2679

\bibitem[{{Borucki} {et~al.}(2010){Borucki}, {Koch}, {Basri}, {Batalha}, {Brown}, {Caldwell}, {Caldwell}, {Christensen-Dalsgaard}, {Cochran}, {DeVore}, {Dunham}, {Dupree}, {Gautier}, {Geary}, {Gilliland}, {Gould}, {Howell}, {Jenkins}, {Kondo}, {Latham}, {Marcy}, {Meibom}, {Kjeldsen}, {Lissauer}, {Monet}, {Morrison}, {Sasselov}, {Tarter}, {Boss}, {Brownlee}, {Owen}, {Buzasi}, {Charbonneau}, {Doyle}, {Fortney}, {Ford}, {Holman}, {Seager}, {Steffen}, {Welsh}, {Rowe}, {Anderson}, {Buchhave}, {Ciardi}, {Walkowicz}, {Sherry}, {Horch}, {Isaacson}, {Everett}, {Fischer}, {Torres}, {Johnson}, {Endl}, {MacQueen}, {Bryson}, {Dotson}, {Haas}, {Kolodziejczak}, {Van Cleve}, {Chandrasekaran}, {Twicken}, {Quintana}, {Clarke}, {Allen}, {Li}, {Wu}, {Tenenbaum}, {Verner}, {Bruhweiler}, {Barnes}, \& {Prsa}}]{Borucki_2010}
{Borucki}, W.~J., {Koch}, D., {Basri}, G., {et~al.} 2010, Science, 327, 977

\bibitem[{{Bradley} {et~al.}(2022){Bradley}, {Sip{\H{o}}cz}, {Robitaille}, {Tollerud}, {Vin{\'\i}cius}, {Deil}, {Barbary}, {Wilson}, {Busko}, {Donath}, {G{\"u}nther}, {Cara}, {Lim}, {Me{\ss}linger}, {Conseil}, {Bostroem}, {Droettboom}, {Bray}, {Andersen Bratholm}, {Barentsen}, {Craig}, {Rathi}, {Pascual}, {Perren}, {Georgiev}, {De Val-Borro}, {Kerzendorf}, {Bach}, {Quint}, \& {Souchereau}}]{Bradley_2022}
{Bradley}, L., {Sip{\H{o}}cz}, B., {Robitaille}, T., {et~al.} 2022, {astropy/photutils: 1.5.0}

\bibitem[{{Burrows}(2014)}]{Burrows_2014}
{Burrows}, A.~S. 2014, PNAS, 111, 12601

\bibitem[{Castan \& Menou(2011)}]{Castan_2011}
Castan, T. \& Menou, K. 2011, \apjl, 743, L36

\bibitem[{{Chaturvedi} {et~al.}(2022){Chaturvedi}, {Bluhm, P.}, {Nagel, E.}, {Hatzes, A. P.}, {Morello, G.}, {Brady, M.}, {Korth, J.}, {Molaverdikhani, K.}, {Kossakowski, D.}, {Caballero, J. A.}, {Guenther, E. W.}, {Pallé, E.}, {Espinoza, N.}, {Seifahrt, A.}, {Lodieu, N.}, {Cifuentes, C.}, {Furlan, E.}, {Amado, P. J.}, {Barclay, T.}, {Bean, J.}, {Béjar, V. J. S.}, {Bergond, G.}, {Boyle, A. W.}, {Ciardi, D.}, {Collins, K. A.}, {Collins, K. I.}, {Esparza-Borges, E.}, {Fukui, A.}, {Gnilka, C. L.}, {Goeke, R.}, {Guerra, P.}, {Henning, Th.}, {Herrero, E.}, {Howell, S. B.}, {Jeffers, S. V.}, {Jenkins, J. M.}, {Jensen, E. L. N.}, {Kasper, D.}, {Kodama, T.}, {Latham, D. W.}, {López-González, M. J.}, {Luque, R.}, {Montes, D.}, {Morales, J. C.}, {Mori, M.}, {Murgas, F.}, {Narita, N.}, {Nowak, G.}, {Parviainen, H.}, {Passegger, V. M.}, {Quirrenbach, A.}, {Reffert, S.}, {Reiners, A.}, {Ribas, I.}, {Ricker, G. R.}, {Rodriguez, E.}, {Rodríguez-López, C.}, {Schlecker, M.}, {Schwarz, R. P.}, {Schweitzer, A.}, {Seager,
  S.}, {Stefánsson, G.}, {Stockdale, C.}, {Tal-Or, L.}, {Twicken, J. D.}, {Vanaverbeke, S.}, {Wang, G.}, {Watanabe, D.}, {Winn, J. N.}, \& {Zechmeister, M.}}]{Chaturvedi_2022}
{Chaturvedi}, P., {Bluhm, P.}, {Nagel, E.}, {et~al.} 2022, \aap, 666, A155

\bibitem[{{Correia} \& {Laskar}(2010)}]{Correia_Laskar_2010}
{Correia}, A.~C.~M. \& {Laskar}, J. 2010, in Exoplanets, ed. S.~{Seager}, 239--266

\bibitem[{Cowan \& Agol(2010)}]{Cowan_2011}
Cowan, N.~B. \& Agol, E. 2010, \apj, 726, 82

\bibitem[{{Cox}(2000)}]{Cox_2000}
{Cox}, A.~N. 2000, {Allen's astrophysical quantities}

\bibitem[{{Crossfield} {et~al.}(2022){Crossfield}, {Malik}, {Hill}, {Kane}, {Foley}, {Polanski}, {Coria}, {Brande}, {Zhang}, {Wienke}, {Kreidberg}, {Cowan}, {Dragomir}, {Gorjian}, {Mikal-Evans}, {Benneke}, {Christiansen}, {Deming}, \& {Morales}}]{Crossfield_2022}
{Crossfield}, I. J.~M., {Malik}, M., {Hill}, M.~L., {et~al.} 2022, \apjl, 937, L17

\bibitem[{Crossfield {et~al.}(2019)Crossfield, Waalkes, Newton, Narita, Muirhead, Ment, Matthews, Kraus, Kostov, Kosiarek, Kane, Isaacson, Halverson, Gonzales, Everett, Dragomir, Collins, Chontos, Berardo, Winters, Winn, Scott, Rojas-Ayala, Rizzuto, Petigura, Peterson, Mocnik, Mikal-Evans, Mehrle, Matson, Kuzuhara, Irwin, Huber, Huang, Howell, Howard, Hirano, Fulton, Dupuy, Dressing, Dalba, Charbonneau, Burt, Berta-Thompson, Benneke, Watanabe, Twicken, Tamura, Schlieder, Seager, Rose, Ricker, Quintana, L{\'e}pine, Latham, Kotani, Jenkins, Hori, Colon, \& Caldwell}]{Crossfield_2019}
Crossfield, I. J.~M., Waalkes, W., Newton, E.~R., {et~al.} 2019, \apjl, 883, L16

\bibitem[{{Damiano} {et~al.}(2022){Damiano}, {Hu}, {Barclay}, {Zieba}, {Kreidberg}, {Brande}, {Colon}, {Covone}, {Crossfield}, {Domagal-Goldman}, {Fauchez}, {Fiscale}, {Gallo}, {Gilbert}, {Hedges}, {Kite}, {Kopparapu}, {Kostov}, {Morley}, {Mullally}, {Pidhorodetska}, {Schlieder}, \& {Quintana}}]{Damiano_2022}
{Damiano}, M., {Hu}, R., {Barclay}, T., {et~al.} 2022, \aj, 164, 225

\bibitem[{{de Wit} {et~al.}(2018){de Wit}, {Wakeford}, {Lewis}, {Delrez}, {Gillon}, {Selsis}, {Leconte}, {Demory}, {Bolmont}, {Bourrier}, {Burgasser}, {Grimm}, {Jehin}, {Lederer}, {Owen}, {Stamenkovi{\'c}}, \& {Triaud}}]{deWit_2018}
{de Wit}, J., {Wakeford}, H.~R., {Lewis}, N.~K., {et~al.} 2018, Nature Astronomy, 2, 214

\bibitem[{Delrez {et~al.}(2018)Delrez, Gillon, Queloz, Demory, Almleaky, de~Wit, Jehin, Triaud, Barkaoui, Burdanov, Burgasser, Ducrot, McCormac, Murray, Fernandes, Sohy, Thompson, Grootel, Alonso, Benkhaldoun, \& Rebolo}]{Delrez_2018}
Delrez, L., Gillon, M., Queloz, D., {et~al.} 2018, in Ground-based and Airborne Telescopes VII, ed. H.~K. Marshall \& J.~Spyromilio, Vol. 10700, International Society for Optics and Photonics (SPIE), 446 -- 466

\bibitem[{{Diamond-Lowe} {et~al.}(2018){Diamond-Lowe}, {Berta-Thompson}, {Charbonneau}, \& {Kempton}}]{Diamond_2018}
{Diamond-Lowe}, H., {Berta-Thompson}, Z., {Charbonneau}, D., \& {Kempton}, E. M.~R. 2018, \aj, 156, 42

\bibitem[{{Diamond-Lowe} {et~al.}(2020){Diamond-Lowe}, {Charbonneau}, {Malik}, {Kempton}, \& {Beletsky}}]{Diamond_2020}
{Diamond-Lowe}, H., {Charbonneau}, D., {Malik}, M., {Kempton}, E. M.~R., \& {Beletsky}, Y. 2020, \aj, 160, 188

\bibitem[{{Diamond-Lowe} {et~al.}(2023{\natexlab{a}}){Diamond-Lowe}, {Mendonca}, {Akin}, {Allen}, {Baungaard}, {Borsato}, {Buchhave}, {Burgasser}, {Demory}, {Espinoza}, {Fisher}, {Fortune}, {Gibson}, {Gressier}, {Guzman Mesa}, {Heng}, {Hoeijmakers}, {Hooton}, {Jones}, {Kitzmann}, {Lueber}, {Meier Valdes}, {Prinoth}, {Rathcke}, \& {Tian}}]{Diamond_lowe_2023}
{Diamond-Lowe}, H., {Mendonca}, J.~M., {Akin}, C.~J., {et~al.} 2023{\natexlab{a}}, {The Hot Rocks Survey: Testing 9 Irradiated Terrestrial Exoplanets for Atmospheres}, JWST Proposal. Cycle 2, ID. \#3730

\bibitem[{{Diamond-Lowe} {et~al.}(2023{\natexlab{b}}){Diamond-Lowe}, {Mendon{\c{c}}a}, {Charbonneau}, \& {Buchhave}}]{Diamond_2022}
{Diamond-Lowe}, H., {Mendon{\c{c}}a}, J.~M., {Charbonneau}, D., \& {Buchhave}, L.~A. 2023{\natexlab{b}}, \aj, 165, 169

\bibitem[{{Diamond-Lowe} {et~al.}(2021){Diamond-Lowe}, {Youngblood}, {Charbonneau}, {King}, {Teal}, {Bastelberger}, {Corrales}, \& {Kempton}}]{Diamond_2021}
{Diamond-Lowe}, H., {Youngblood}, A., {Charbonneau}, D., {et~al.} 2021, \aj, 162, 10

\bibitem[{{Dietrich} {et~al.}(2023){Dietrich}, {Apai}, {Schlecker}, {Hardegree-Ullman}, {Rackham}, {Kurtovic}, {Molaverdikhani}, {Gabor}, {Henning}, {Chen}, {Mancini}, {Bixel}, {Gibbs}, {Boyle}, {Brown-Sevilla}, {Burn}, {Delage}, {Flores-Rivera}, {Franceschi}, {Pichierri}, {Savvidou}, {Syed}, {Bruni}, {Ip}, {Ngeow}, {Tsai}, {Lin}, {Hou}, {Hsiao}, {Lin}, {Lin}, {Basant}, \& {EDEN Project}}]{Dietrich_2023}
{Dietrich}, J., {Apai}, D., {Schlecker}, M., {et~al.} 2023, \aj, 165, 149

\bibitem[{{Donati} {et~al.}(2020){Donati}, {Kouach}, {Moutou}, {Doyon}, {Delfosse}, {Artigau}, {Baratchart}, {Lacombe}, {Barrick}, {H{\'e}brard}, {Bouchy}, {Saddlemyer}, {Par{\`e}s}, {Rabou}, {Micheau}, {Dolon}, {Reshetov}, {Challita}, {Carmona}, {Striebig}, {Thibault}, {Martioli}, {Cook}, {Fouqu{\'e}}, {Vermeulen}, {Wang}, {Arnold}, {Pepe}, {Boisse}, {Figueira}, {Bouvier}, {Ray}, {Feugeade}, {Morin}, {Alencar}, {Hobson}, {Castilho}, {Udry}, {Santos}, {Hernandez}, {Benedict}, {Vall{\'e}e}, {Gallou}, {Dupieux}, {Larrieu}, {Perruchot}, {Sottile}, {Moreau}, {Usher}, {Baril}, {Wildi}, {Chazelas}, {Malo}, {Bonfils}, {Loop}, {Kerley}, {Wevers}, {Dunn}, {Pazder}, {Macdonald}, {Dubois}, {Carri{\'e}}, {Valentin}, {Henault}, {Yan}, \& {Steinmetz}}]{Donati_2020}
{Donati}, J.~F., {Kouach}, D., {Moutou}, C., {et~al.} 2020, \mnras, 498, 5684

\bibitem[{Dressing \& Charbonneau(2015)}]{Dressing_2015}
Dressing, C.~D. \& Charbonneau, D. 2015, \apj, 807, 45

\bibitem[{{Driscoll} \& {Barnes}(2015)}]{Driscoll_2015}
{Driscoll}, P.~E. \& {Barnes}, R. 2015, Astrobiology, 15, 739

\bibitem[{Ducrot {et~al.}(2025)Ducrot, Lagage, Min, Gillon, Bell, Tremblin, Greene, Dyrek, Bouwman, Waters, G{\"u}del, Henning, Vandenbussche, Absil, Barrado, Boccaletti, Coulais, Decin, Edwards, Gastaud, Glasse, Kendrew, Olofsson, Patapis, Pye, Rouan, Whiteford, Argyriou, Cossou, Glauser, Krause, Lahuis, Royer, Scheithauer, Colina, van Dishoeck, Ostlin, Ray, \& Wright}]{Ducrot_2024}
Ducrot, E., Lagage, P.-O., Min, M., {et~al.} 2025, Nature Astronomy, 9, 358

\bibitem[{{Dyrek} {et~al.}(2024){Dyrek}, {Ducrot}, {Lagage}, {Tremblin}, {Kendrew}, {Bouwman}, \& {Bouffet}}]{Dyrek_2024}
{Dyrek}, A., {Ducrot}, E., {Lagage}, P.~O., {et~al.} 2024, \aap, 683, A212

\bibitem[{{Eastman} {et~al.}(2019){Eastman}, {Rodriguez}, {Agol}, {Stassun}, {Beatty}, {Vanderburg}, {Gaudi}, {Collins}, \& {Luger}}]{Eastman_2019}
{Eastman}, J.~D., {Rodriguez}, J.~E., {Agol}, E., {et~al.} 2019, \pasp, arXiv:1907.09480, submitted

\bibitem[{Eschen \& Kunimoto(2024)}]{Eschen_2024}
Eschen, Y. N.~E. \& Kunimoto, M. 2024, \mnras, 531, 5053

\bibitem[{Essack {et~al.}(2020)Essack, Seager, \& Pajusalu}]{Essack_2020}
Essack, Z., Seager, S., \& Pajusalu, M. 2020, \apj, 898, 160

\bibitem[{{Fauchez} {et~al.}(2025){Fauchez}, {Rackham}, {Ducrot}, {Stevenson}, \& {de Wit}}]{Fauchez_2025}
{Fauchez}, T.~J., {Rackham}, B.~V., {Ducrot}, E., {Stevenson}, K.~B., \& {de Wit}, J. 2025, \apjl, arXiv:2502.19585, submitted

\bibitem[{Foreman-Mackey(2016)}]{corner_2016}
Foreman-Mackey, D. 2016, JOSS, 1, 24

\bibitem[{Foreman-Mackey {et~al.}(2021)Foreman-Mackey, Savel, Luger, Agol, Czekala, Price-Whelan, Hedges, Gilbert, Bouma, Barclay, \& Brandt}]{Foreman-Mackey_2021}
Foreman-Mackey, D., Savel, A., Luger, R., {et~al.} 2021, exoplanet-dev/exoplanet v0.5.0

\bibitem[{{Fortune} {et~al.}(2025){Fortune}, {Gibson}, \& {Diamond-Lowe}}]{Fortune_2025}
{Fortune}, M., {Gibson}, N.~P., \& {Diamond-Lowe}, H. 2025, \aap

\bibitem[{{Fulton} {et~al.}(2017){Fulton}, {Petigura}, {Howard}, {Isaacson}, {Marcy}, {Cargile}, {Hebb}, {Weiss}, {Johnson}, {Morton}, {Sinukoff}, {Crossfield}, \& {Hirsch}}]{Fulton_2017}
{Fulton}, B.~J., {Petigura}, E.~A., {Howard}, A.~W., {et~al.} 2017, \aj, 154, 109

\bibitem[{{Garraffo} {et~al.}(2016){Garraffo}, {Drake}, \& {Cohen}}]{Garraffo_2016}
{Garraffo}, C., {Drake}, J.~J., \& {Cohen}, O. 2016, \apjl, 833, L4

\bibitem[{{Garraffo} {et~al.}(2017){Garraffo}, {Drake}, {Cohen}, {Alvarado-G{\'o}mez}, \& {Moschou}}]{Garraffo_2017}
{Garraffo}, C., {Drake}, J.~J., {Cohen}, O., {Alvarado-G{\'o}mez}, J.~D., \& {Moschou}, S.~P. 2017, \apjl, 843, L33

\bibitem[{Gelman \& Rubin(1992)}]{Gelman_1992}
Gelman, A. \& Rubin, D.~B. 1992, Statistical Science, 7, 457

\bibitem[{{Gibbs} {et~al.}(2020){Gibbs}, {Bixel}, {Rackham}, {Apai}, {Schlecker}, {Espinoza}, {Mancini}, {Chen}, {Henning}, {Gabor}, {Boyle}, {Perez Chavez}, {Mousseau}, {Dietrich}, {Jay Socia}, {Ip}, {Ngeow}, {Tsai}, {Bhandare}, {Marian}, {Baehr}, {Brown}, {H{\"a}berle}, {Keppler}, {Molaverdikhani}, \& {Sarkis}}]{Gibbs_2020}
{Gibbs}, A., {Bixel}, A., {Rackham}, B.~V., {et~al.} 2020, \aj, 159, 169

\bibitem[{{Gilbert} {et~al.}(2020){Gilbert}, {Barclay}, {Schlieder}, {Quintana}, {Hord}, {Kostov}, {Lopez}, {Rowe}, {Hoffman}, {Walkowicz}, {Silverstein}, {Rodriguez}, {Vanderburg}, {Suissa}, {Airapetian}, {Clement}, {Raymond}, {Mann}, {Kruse}, {Lissauer}, {Col{\'o}n}, {Kopparapu}, {Kreidberg}, {Zieba}, {Collins}, {Quinn}, {Howell}, {Ziegler}, {Vrijmoet}, {Adams}, {Arney}, {Boyd}, {Brande}, {Burke}, {Cacciapuoti}, {Chance}, {Christiansen}, {Covone}, {Daylan}, {Dineen}, {Dressing}, {Essack}, {Fauchez}, {Galgano}, {Howe}, {Kaltenegger}, {Kane}, {Lam}, {Lee}, {Lewis}, {Logsdon}, {Mandell}, {Monsue}, {Mullally}, {Mullally}, {Paudel}, {Pidhorodetska}, {Plavchan}, {Reyes}, {Rinehart}, {Rojas-Ayala}, {Smith}, {Stassun}, {Tenenbaum}, {Vega}, {Villanueva}, {Wolf}, {Youngblood}, {Ricker}, {Vanderspek}, {Latham}, {Seager}, {Winn}, {Jenkins}, {Bakos}, {Brice{\~n}o}, {Ciardi}, {Cloutier}, {Conti}, {Couperus}, {Di Sora}, {Eisner}, {Everett}, {Gan}, {Hartman}, {Henry}, {Isopi}, {Jao}, {Jensen}, {Law}, {Mallia}, {Matson},
  {Shappee}, {Le Wood}, \& {Winters}}]{Gilbert_2020}
{Gilbert}, E.~A., {Barclay}, T., {Schlieder}, J.~E., {et~al.} 2020, \aj, 160, 116

\bibitem[{{Gillon} {et~al.}(2016){Gillon}, {Jehin}, {Lederer}, {Delrez}, {de Wit}, {Burdanov}, {Van Grootel}, {Burgasser}, {Triaud}, {Opitom}, {Demory}, {Sahu}, {Bardalez Gagliuffi}, {Magain}, \& {Queloz}}]{Gillon_2016}
{Gillon}, M., {Jehin}, E., {Lederer}, S.~M., {et~al.} 2016, \nat, 533, 221

\bibitem[{{Gillon} {et~al.}(2013){Gillon}, {Jehin, E.}, {Fumel, A.}, {Magain, P.}, \& {Queloz, D.}}]{Gillon_2013}
{Gillon}, M., {Jehin, E.}, {Fumel, A.}, {Magain, P.}, \& {Queloz, D.} 2013, EPJ Web of Conferences, 47, 03001

\bibitem[{Gillon {et~al.}(2024)Gillon, Pedersen, Rackham, Dransfield, Ducrot, Barkaoui, Burdanov, Schroffenegger, G{\'o}mez Maqueo~Chew, Lederer, Alonso, Burgasser, Howell, Narita, de~Wit, Demory, Queloz, Triaud, Delrez, Jehin, Hooton, Garcia, Jano~Mu{\~n}oz, Murray, Pozuelos, Sebastian, Timmermans, Thompson, Z{\'u}{\~n}iga-Fern{\'a}ndez, Aceituno, Aganze, Amado, Baycroft, Benkhaldoun, Berardo, Bolmont, Clark, Davis, Davoudi, de~Beurs, de~Leon, Ikoma, Ikuta, Isogai, Fukuda, Fukui, Gerasimov, Ghachoui, G{\"u}nther, Hasler, Hayashi, Heng, Hu, Kagetani, Kawai, Kawauchi, Kitzmann, Koll, Lendl, Livingston, Lyu, Meier~Vald{\'e}s, Mori, McCormac, Murgas, Niraula, Pall{\'e}, Plauchu-Frayn, Rebolo, Sabin, Schackey, Schanche, Selsis, Sota, Stalport, Standing, Stassun, Tamura, Terada, Theissen, Turbet, Van~Grootel, Varas, Watanabe, \& Zong~Lang}]{Gillon_2024}
Gillon, M., Pedersen, P.~P., Rackham, B.~V., {et~al.} 2024, Nature Astronomy

\bibitem[{Ginsburg {et~al.}(2019)Ginsburg, Sip{\H{o}}cz, Brasseur, Cowperthwaite, Craig, Deil, Guillochon, Guzman, Liedtke, Lim, Lockhart, Mommert, Morris, Norman, Parikh, Persson, Robitaille, Segovia, Singer, Tollerud, de~Val-Borro, Valtchanov, \& and}]{Ginsburg_2019}
Ginsburg, A., Sip{\H{o}}cz, B.~M., Brasseur, C.~E., {et~al.} 2019, \aj, 157, 98

\bibitem[{{Gordon} {et~al.}(2025){Gordon}, {Sloan}, {Garcia Marin}, {Libralato}, {Rieke}, {Aguilar}, {Bohlin}, {Cracraft}, {Decleir}, {Gaspar}, {Kendrew}, {Law}, {Noriega-Crespo}, \& {Regan}}]{Gordon_2025}
{Gordon}, K.~D., {Sloan}, G.~C., {Garcia Marin}, M., {et~al.} 2025, \aj, 169, 6

\bibitem[{{Gould} {et~al.}(2003){Gould}, {Pepper}, \& {DePoy}}]{Gould_2003}
{Gould}, A., {Pepper}, J., \& {DePoy}, D.~L. 2003, \apj, 594, 533

\bibitem[{{Greene} {et~al.}(2023){Greene}, {Bell}, {Ducrot}, {Dyrek}, {Lagage}, \& {Fortney}}]{Greene_2023}
{Greene}, T.~P., {Bell}, T.~J., {Ducrot}, E., {et~al.} 2023, \nat, 618, 39

\bibitem[{{Gressier} {et~al.}(2024){Gressier}, {Espinoza}, {Allen}, {Sing}, {Banerjee}, {Barstow}, {Valenti}, {Lewis}, {Birkmann}, {Challener}, {Manjavacas}, {Alves de Oliveira}, {Crouzet}, \& {Beck}}]{Gressier_2024}
{Gressier}, A., {Espinoza}, N., {Allen}, N.~H., {et~al.} 2024, \apjl, 975, L10

\bibitem[{{Grimm} {et~al.}(2021){Grimm}, {Malik}, {Kitzmann}, {Guzm{\'a}n-Mesa}, {Hoeijmakers}, {Fisher}, {Mendon{\c{c}}a}, {Yurchenko}, {Tennyson}, {Alesina}, {Buchschacher}, {Burnier}, {Segransan}, {Kurucz}, \& {Heng}}]{Grimm_2021}
{Grimm}, S.~L., {Malik}, M., {Kitzmann}, D., {et~al.} 2021, \apjs, 253, 30

\bibitem[{G{\"u}nther {et~al.}(2019)G{\"u}nther, Pozuelos, Dittmann, Dragomir, Kane, Daylan, Feinstein, Huang, Morton, Bonfanti, Bouma, Burt, Collins, Lissauer, Matthews, Montet, Vanderburg, Wang, Winters, Ricker, Vanderspek, Latham, Seager, Winn, Jenkins, Armstrong, Barkaoui, Batalha, Bean, Caldwell, Ciardi, Collins, Crossfield, Fausnaugh, Furesz, Gan, Gillon, Guerrero, Horne, Howell, Ireland, Isopi, Jehin, Kielkopf, Lepine, Mallia, Matson, Myers, Palle, Quinn, Relles, Rojas-Ayala, Schlieder, Sefako, Shporer, Su{\'a}rez, Tan, Ting, Twicken, \& Waite}]{Gunther_2019}
G{\"u}nther, M.~N., Pozuelos, F.~J., Dittmann, J.~A., {et~al.} 2019, Nature Astronomy, 3, 1099

\bibitem[{{Hammond} {et~al.}(2025){Hammond}, {Guimond}, {Lichtenberg}, {Nicholls}, {Fisher}, {Luque}, {Meier}, {Taylor}, {Changeat}, {Dang}, {Hay}, {Herbort}, \& {Teske}}]{Hammond_2024}
{Hammond}, M., {Guimond}, C.~M., {Lichtenberg}, T., {et~al.} 2025, \apjl, 978, L40

\bibitem[{Hansen(2008)}]{Hansen_2008}
Hansen, B. M.~S. 2008, \apjs, 179, 484

\bibitem[{Harris {et~al.}(2020)Harris, Millman, van~der Walt, Gommers, Virtanen, Cournapeau, Wieser, Taylor, Berg, Smith, Kern, Picus, Hoyer, van Kerkwijk, Brett, Haldane, del R{\'{i}}o, Wiebe, Peterson, G{\'{e}}rard-Marchant, Sheppard, Reddy, Weckesser, Abbasi, Gohlke, \& Oliphant}]{Harris_2020}
Harris, C.~R., Millman, K.~J., van~der Walt, S.~J., {et~al.} 2020, Nature, 585, 357

\bibitem[{Hoffman \& Gelman(2011)}]{Hoffmann_2011}
Hoffman, M.~D. \& Gelman, A. 2011, The No-U-Turn Sampler: Adaptively Setting Path Lengths in Hamiltonian Monte Carlo

\bibitem[{{Hu} {et~al.}(2024){Hu}, {Bello-Arufe}, {Zhang}, {Paragas}, {Zilinskas}, {van Buchem}, {Bess}, {Patel}, {Ito}, {Damiano}, {Scheucher}, {Oza}, {Knutson}, {Miguel}, {Dragomir}, {Brandeker}, \& {Demory}}]{Hu_2024}
{Hu}, R., {Bello-Arufe}, A., {Zhang}, M., {et~al.} 2024, \nat, 630, 609

\bibitem[{Hunter(2007)}]{Hunter_2007}
Hunter, J.~D. 2007, Computing in Science \& Engineering, 9, 90

\bibitem[{{Ih} {et~al.}(2023){Ih}, {Kempton}, {Whittaker}, \& {Lessard}}]{Ih_2023}
{Ih}, J., {Kempton}, E. M.~R., {Whittaker}, E.~A., \& {Lessard}, M. 2023, \apjl, 952, L4

\bibitem[{{Jehin} {et~al.}(2018){Jehin}, {Gillon}, {Queloz}, {Delrez}, {Burdanov}, {Murray}, {Sohy}, {Ducrot}, {Sebastian}, {Thompson}, {McCormac}, {Almleaky}, {Burgasser}, {Demory}, {de Wit}, {Barkaoui}, {Pozuelos}, {Triaud}, \& {Van Grootel}}]{Jehin_2018}
{Jehin}, E., {Gillon}, M., {Queloz}, D., {et~al.} 2018, The Messenger, 174, 2

\bibitem[{{Kipping}(2013)}]{exoplanet:kipping13}
{Kipping}, D.~M. 2013, \mnras, 435, 2152

\bibitem[{{Kislyakova} {et~al.}(2018){Kislyakova}, {Fossati}, {Johnstone}, {Noack}, {L{\"u}ftinger}, {Zaitsev}, \& {Lammer}}]{Kislyakova_2018}
{Kislyakova}, K.~G., {Fossati}, L., {Johnstone}, C.~P., {et~al.} 2018, \apj, 858, 105

\bibitem[{{Kitzmann} {et~al.}(2020){Kitzmann}, {Heng}, {Oreshenko}, {Grimm}, {Apai}, {Bowler}, {Burgasser}, \& {Marley}}]{Kitzmann_2020}
{Kitzmann}, D., {Heng}, K., {Oreshenko}, M., {et~al.} 2020, \apj, 890, 174

\bibitem[{{Koll}(2022)}]{Koll_2022}
{Koll}, D. D.~B. 2022, \apj, 924, 134

\bibitem[{Kreidberg(2015)}]{Kreidberg_2015}
Kreidberg, L. 2015, \pasp, 127, 1161

\bibitem[{Kreidberg {et~al.}(2019)Kreidberg, Koll, Morley, Hu, Schaefer, Deming, Stevenson, Dittmann, Vanderburg, Berardo, Guo, Stassun, Crossfield, Charbonneau, Latham, Loeb, Ricker, Seager, \& Vanderspek}]{Kreidberg_2019}
Kreidberg, L., Koll, D. D.~B., Morley, C., {et~al.} 2019, Nature, 573, 87

\bibitem[{Kumar {et~al.}(2019)Kumar, Carroll, Hartikainen, \& Martin}]{exoplanet:arviz}
Kumar, R., Carroll, C., Hartikainen, A., \& Martin, O.~A. 2019, JOSS

\bibitem[{{Libby-Roberts} {et~al.}(2022){Libby-Roberts}, {Berta-Thompson}, {Diamond-Lowe}, {Gully-Santiago}, {Irwin}, {Kempton}, {Rackham}, {Charbonneau}, {D{\'e}sert}, {Dittmann}, {Hofmann}, {Morley}, \& {Newton}}]{Libby_2022}
{Libby-Roberts}, J.~E., {Berta-Thompson}, Z.~K., {Diamond-Lowe}, H., {et~al.} 2022, \aj, 164, 59

\bibitem[{{Libralato} {et~al.}(2024){Libralato}, {Argyriou}, {Dicken}, {Garc{\'\i}a Mar{\'\i}n}, {Guillard}, {Hines}, {Kavanagh}, {Kendrew}, {Law}, {Noriega-Crespo}, \& {{\'A}lvarez-M{\'a}rquez}}]{Libralato_2024}
{Libralato}, M., {Argyriou}, I., {Dicken}, D., {et~al.} 2024, \pasp, 136, 034502

\bibitem[{{Lim} {et~al.}(2023){Lim}, {Benneke}, {Doyon}, {MacDonald}, {Piaulet}, {Artigau}, {Coulombe}, {Radica}, {L'Heureux}, {Albert}, {Rackham}, {de Wit}, {Salhi}, {Roy}, {Flagg}, {Fournier-Tondreau}, {Taylor}, {Cook}, {Lafreni{\`e}re}, {Cowan}, {Kaltenegger}, {Rowe}, {Espinoza}, {Dang}, \& {Darveau-Bernier}}]{Lim_2023}
{Lim}, O., {Benneke}, B., {Doyon}, R., {et~al.} 2023, \apjl, 955, L22

\bibitem[{{Loyd} {et~al.}(2018){Loyd}, {France}, {Youngblood}, {Schneider}, {Brown}, {Hu}, {Segura}, {Linsky}, {Redfield}, {Tian}, {Rugheimer}, {Miguel}, \& {Froning}}]{Loyd_2018}
{Loyd}, R.~O.~P., {France}, K., {Youngblood}, A., {et~al.} 2018, \apj, 867, 71

\bibitem[{{Luger} {et~al.}(2019){Luger}, {Agol}, {Foreman-Mackey}, {Fleming}, {Lustig-Yaeger}, \& {Deitrick}}]{exoplanet:luger18}
{Luger}, R., {Agol}, E., {Foreman-Mackey}, D., {et~al.} 2019, \aj, 157, 64

\bibitem[{{Lustig-Yaeger} {et~al.}(2023){Lustig-Yaeger}, {Fu}, {May}, {Ceballos}, {Moran}, {Peacock}, {Stevenson}, {Kirk}, {L{\'o}pez-Morales}, {MacDonald}, {Mayorga}, {Sing}, {Sotzen}, {Valenti}, {Redai}, {Alam}, {Batalha}, {Bennett}, {Gonzalez-Quiles}, {Kruse}, {Lothringer}, {Rustamkulov}, \& {Wakeford}}]{Lustig_2023}
{Lustig-Yaeger}, J., {Fu}, G., {May}, E.~M., {et~al.} 2023, Nature Astronomy, 7, 1317

\bibitem[{{Malik} {et~al.}(2017){Malik}, {Grosheintz}, {Mendon{\c{c}}a}, {Grimm}, {Lavie}, {Kitzmann}, {Tsai}, {Burrows}, {Kreidberg}, {Bedell}, {Bean}, {Stevenson}, \& {Heng}}]{Malik_2017}
{Malik}, M., {Grosheintz}, L., {Mendon{\c{c}}a}, J.~M., {et~al.} 2017, \aj, 153, 56

\bibitem[{{Malik} {et~al.}(2019{\natexlab{a}}){Malik}, {Kempton}, {Koll}, {Mansfield}, {Bean}, \& {Kite}}]{Malik_2019b}
{Malik}, M., {Kempton}, E. M.~R., {Koll}, D. D.~B., {et~al.} 2019{\natexlab{a}}, \apj, 886, 142

\bibitem[{{Malik} {et~al.}(2019{\natexlab{b}}){Malik}, {Kitzmann}, {Mendon{\c{c}}a}, {Grimm}, {Marleau}, {Linder}, {Tsai}, \& {Heng}}]{Malik_2019a}
{Malik}, M., {Kitzmann}, D., {Mendon{\c{c}}a}, J.~M., {et~al.} 2019{\natexlab{b}}, \aj, 157, 170

\bibitem[{{Mallonn} {et~al.}(2019){Mallonn}, {K{\"o}hler}, {Alexoudi}, {von Essen}, {Granzer}, {Poppenhaeger}, \& {Strassmeier}}]{Mallonn_2019}
{Mallonn}, M., {K{\"o}hler}, J., {Alexoudi}, X., {et~al.} 2019, \aap, 624, A62

\bibitem[{Mandel \& Agol(2002)}]{Mandel_2002}
Mandel, K. \& Agol, E. 2002, \apj, 580, L171

\bibitem[{{Marley} {et~al.}(2013){Marley}, {Ackerman}, {Cuzzi}, \& {Kitzmann}}]{Marley_2013}
{Marley}, M.~S., {Ackerman}, A.~S., {Cuzzi}, J.~N., \& {Kitzmann}, D. 2013, in Comparative Climatology of Terrestrial Planets, ed. S.~J. {Mackwell}, A.~A. {Simon-Miller}, J.~W. {Harder}, \& M.~A. {Bullock}, 367--392

\bibitem[{{May} {et~al.}(2023){May}, {MacDonald}, {Bennett}, {Moran}, {Wakeford}, {Peacock}, {Lustig-Yaeger}, {Highland}, {Stevenson}, {Sing}, {Mayorga}, {Batalha}, {Kirk}, {L{\'o}pez-Morales}, {Valenti}, {Alam}, {Alderson}, {Fu}, {Gonzalez-Quiles}, {Lothringer}, {Rustamkulov}, \& {Sotzen}}]{May_2023}
{May}, E.~M., {MacDonald}, R.~J., {Bennett}, K.~A., {et~al.} 2023, \apjl, 959, L9

\bibitem[{McElreath(2016)}]{McElreath_2016}
McElreath, R. 2016, Statistical Rethinking: A Bayesian Course with Examples in R and Stan, Chapman \& Hall/CRC Texts in Statistical Science (CRC Press)

\bibitem[{{Meier} {et~al.}(2021){Meier}, {Bower}, {Lichtenberg}, {Tackley}, \& {Demory}}]{Tobias_2021}
{Meier}, T.~G., {Bower}, D.~J., {Lichtenberg}, T., {Tackley}, P.~J., \& {Demory}, B.-O. 2021, \apjl, 908, L48

\bibitem[{{Meier Vald{\'e}s} {et~al.}(2023){Meier Vald{\'e}s}, {Morris}, {Demory}, {Brandeker}, {Kitzmann}, {Benz}, {Deline}, {Flor{\'e}n}, {Sousa}, {Bourrier}, {Singh}, {Heng}, {Strugarek}, {Bower}, {J{\"a}ggi}, {Carone}, {Lendl}, {Jones}, {Oza}, {Demangeon}, {Alibert}, {Alonso}, {Anglada}, {Asquier}, {B{\'a}rczy}, {Barrado Navascues}, {Barros}, {Baumjohann}, {Beck}, {Beck}, {Billot}, {Bonfils}, {Borsato}, {Broeg}, {Cabrera}, {Charnoz}, {Collier Cameron}, {Csizmadia}, {Cubillos}, {Davies}, {Deleuil}, {Delrez}, {Ehrenreich}, {Erikson}, {Fortier}, {Fossati}, {Fridlund}, {Gandolfi}, {Gillon}, {G{\"u}del}, {G{\"u}nther}, {Hoyer}, {Isaak}, {Kiss}, {Laskar}, {Lecavelier des Etangs}, {Lovis}, {Magrin}, {Maxted}, {Mordasini}, {Nascimbeni}, {Olofsson}, {Ottensamer}, {Pagano}, {Pall{\'e}}, {Peter}, {Piotto}, {Pollacco}, {Queloz}, {Ragazzoni}, {Rando}, {Rauer}, {Ribas}, {Santos}, {Sarajlic}, {Scandariato}, {S{\'e}gransan}, {Sicilia}, {Simon}, {Smith}, {Steller}, {Szab{\'o}}, {Thomas}, {Udry}, {Ulmer}, {Van Grootel},
  {Venturini}, {Walton}, {Wilson}, \& {Wolter}}]{Meier_2023}
{Meier Vald{\'e}s}, E.~A., {Morris}, B.~M., {Demory}, B.~O., {et~al.} 2023, \aap, 677, A112

\bibitem[{{Meier Vald\'es} {et~al.}(2022){Meier Vald\'es}, {Morris, B. M.}, {Wells, R. D.}, {Schanche, N.}, \& {Demory, B.-O.}}]{Meier_2022}
{Meier Vald\'es}, E.~A., {Morris, B. M.}, {Wells, R. D.}, {Schanche, N.}, \& {Demory, B.-O.} 2022, A\&A, 663, A95

\bibitem[{{Miguel} {et~al.}(2011){Miguel}, {Kaltenegger}, {Fegley}, \& {Schaefer}}]{Miguel_2011}
{Miguel}, Y., {Kaltenegger}, L., {Fegley}, B., \& {Schaefer}, L. 2011, \apjl, 742, L19

\bibitem[{{Moran} {et~al.}(2023){Moran}, {Stevenson}, {Sing}, {MacDonald}, {Kirk}, {Lustig-Yaeger}, {Peacock}, {Mayorga}, {Bennett}, {L{\'o}pez-Morales}, {May}, {Rustamkulov}, {Valenti}, {Adams Redai}, {Alam}, {Batalha}, {Fu}, {Gonzalez-Quiles}, {Highland}, {Kruse}, {Lothringer}, {Ortiz Ceballos}, {Sotzen}, \& {Wakeford}}]{Moran_2023}
{Moran}, S.~E., {Stevenson}, K.~B., {Sing}, D.~K., {et~al.} 2023, \apjl, 948, L11

\bibitem[{{Morrison} {et~al.}(2023){Morrison}, {Dicken}, {Argyriou}, {Ressler}, {Gordon}, {Regan}, {Cracraft}, {Rieke}, {Engesser}, {Alberts}, {Alvarez-Marquez}, {Colbert}, {Fox}, {Gasman}, {Law}, {Garcia Marin}, {G{\'a}sp{\'a}r}, {Guillard}, {Kendrew}, {Labiano}, {Laine}, {Noriega-Crespo}, {Shivaei}, \& {Sloan}}]{Morrison_2023}
{Morrison}, J.~E., {Dicken}, D., {Argyriou}, I., {et~al.} 2023, \pasp, 135, 075004

\bibitem[{{Nutzman} \& {Charbonneau}(2008)}]{Nutzman_2008}
{Nutzman}, P. \& {Charbonneau}, D. 2008, \pasp, 120, 317

\bibitem[{{Owen}(2019)}]{Owen_2019}
{Owen}, J.~E. 2019, Annual Review of Earth and Planetary Sciences, 47, 67

\bibitem[{O’Sullivan \& Aigrain(2024)}]{OSullivan_2024}
O’Sullivan, N.~K. \& Aigrain, S. 2024, \mnras, 531, 4181

\bibitem[{{Park Coy} {et~al.}(2024){Park Coy}, {Ih}, {Kite}, {Koll}, {Tenthoff}, {Bean}, {Weiner Mansfield}, {Zhang}, {Xue}, {Kempton}, {Wolhfarth}, {Hu}, {Lyu}, \& {Wohler}}]{Brandon_2024}
{Park Coy}, B., {Ih}, J., {Kite}, E.~S., {et~al.} 2024, \apj, arXiv:2412.06573, submitted

\bibitem[{{Patel} {et~al.}(2024){Patel}, {Brandeker}, {Kitzmann}, {Petit dit de la Roche}, {Bello-Arufe}, {Heng}, {Meier Vald{\'e}s}, {Persson}, {Zhang}, {Demory}, {Bourrier}, {Deline}, {Ehrenreich}, {Fridlund}, {Hu}, {Lendl}, {Oza}, {Alibert}, \& {Hooton}}]{Patel_2024}
{Patel}, J.~A., {Brandeker}, A., {Kitzmann}, D., {et~al.} 2024, \aap, 690, A159

\bibitem[{{Perryman}(2018)}]{Perryman_2018}
{Perryman}, M. 2018, {The Exoplanet Handbook}

\bibitem[{{Pierrehumbert} \& {Hammond}(2019)}]{Pierrehumbert_2019}
{Pierrehumbert}, R.~T. \& {Hammond}, M. 2019, Annual Review of Fluid Mechanics, 51, 275

\bibitem[{{Pont} {et~al.}(2008){Pont}, {Knutson}, {Gilliland}, {Moutou}, \& {Charbonneau}}]{Pont_2008}
{Pont}, F., {Knutson}, H., {Gilliland}, R.~L., {Moutou}, C., \& {Charbonneau}, D. 2008, \mnras, 385, 109

\bibitem[{Pontoppidan {et~al.}(2016)Pontoppidan, Pickering, Laidler, Gilbert, Sontag, Slocum, Jr., Hanley, Earl, Pueyo, Ravindranath, Karakla, Robberto, Noriega-Crespo, \& Barker}]{Pontoppidan_2016}
Pontoppidan, K.~M., Pickering, T.~E., Laidler, V.~G., {et~al.} 2016, in Observatory Operations: Strategies, Processes, and Systems VI, ed. A.~B. Peck, R.~L. Seaman, \& C.~R. Benn, Vol. 9910, International Society for Optics and Photonics (SPIE), 991016

\bibitem[{Powell {et~al.}(2024)Powell, Feinstein, Lee, Zhang, Tsai, Taylor, Kirk, Bell, Barstow, Gao, Bean, Blecic, Chubb, Crossfield, Jordan, Kitzmann, Moran, Morello, Moses, Welbanks, Yang, Zhang, Ahrer, Bello-Arufe, Brande, Casewell, Crouzet, Cubillos, Demory, Dyrek, Flagg, Hu, Inglis, Jones, Kreidberg, L{\'o}pez-Morales, Lagage, Meier~Vald{\'e}s, Miguel, Parmentier, Piette, Rackham, Radica, Redfield, Stevenson, Wakeford, Aggarwal, Alam, Batalha, Batalha, Benneke, Berta-Thompson, Brady, Caceres, Carter, D{\'e}sert, Harrington, Iro, Line, Lothringer, MacDonald, Mancini, Molaverdikhani, Mukherjee, Nixon, Oza, Palle, Rustamkulov, Sing, Steinrueck, Venot, Wheatley, \& Yurchenko}]{Powell_2024}
Powell, D., Feinstein, A.~D., Lee, E. K.~H., {et~al.} 2024, Nature, 626, 979

\bibitem[{{Quirrenbach} {et~al.}(2014){Quirrenbach}, {Amado}, {Caballero}, {Mundt}, {Reiners}, {Ribas}, {Seifert}, {Abril}, {Aceituno}, {Alonso-Floriano}, {Ammler-von Eiff}, {Antona Jim{\'e}nez}, {Anwand-Heerwart}, {Azzaro}, {Bauer}, {Barrado}, {Becerril}, {B{\'e}jar}, {Ben{\'\i}tez}, {Berdi{\~n}as}, {C{\'a}rdenas}, {Casal}, {Claret}, {Colom{\'e}}, {Cort{\'e}s-Contreras}, {Czesla}, {Doellinger}, {Dreizler}, {Feiz}, {Fern{\'a}ndez}, {Galad{\'\i}}, {G{\'a}lvez-Ortiz}, {Garc{\'\i}a-Piquer}, {Garc{\'\i}a-Vargas}, {Garrido}, {Gesa}, {G{\'o}mez Galera}, {Gonz{\'a}lez {\'A}lvarez}, {Gonz{\'a}lez Hern{\'a}ndez}, {Gr{\"o}zinger}, {Gu{\`a}rdia}, {Guenther}, {de Guindos}, {Guti{\'e}rrez-Soto}, {Hagen}, {Hatzes}, {Hauschildt}, {Helmling}, {Henning}, {Hermann}, {Hern{\'a}ndez Casta{\~n}o}, {Herrero}, {Hidalgo}, {Holgado}, {Huber}, {Huber}, {Jeffers}, {Joergens}, {de Juan}, {Kehr}, {Klein}, {K{\"u}rster}, {Lamert}, {Lalitha}, {Laun}, {Lemke}, {Lenzen}, {L{\'o}pez del Fresno}, {L{\'o}pez Mart{\'\i}}, {L{\'o}pez-Santiago},
  {Mall}, {Mandel}, {Mart{\'\i}n}, {Mart{\'\i}n-Ruiz}, {Mart{\'\i}nez-Rodr{\'\i}guez}, {Marvin}, {Mathar}, {Mirabet}, {Montes}, {Morales Mu{\~n}oz}, {Moya}, {Naranjo}, {Ofir}, {Oreiro}, {Pall{\'e}}, {Panduro}, {Passegger}, {P{\'e}rez-Calpena}, {P{\'e}rez Medialdea}, {Perger}, {Pluto}, {Ram{\'o}n}, {Rebolo}, {Redondo}, {Reffert}, {Reinhardt}, {Rhode}, {Rix}, {Rodler}, {Rodr{\'\i}guez}, {Rodr{\'\i}guez-L{\'o}pez}, {Rodr{\'\i}guez-P{\'e}rez}, {Rohloff}, {Rosich}, {S{\'a}nchez-Blanco}, {S{\'a}nchez Carrasco}, {Sanz-Forcada}, {Sarmiento}, {Sch{\"a}fer}, {Schiller}, {Schmidt}, {Schmitt}, {Solano}, {Stahl}, {Storz}, {St{\"u}rmer}, {Su{\'a}rez}, {Ulbrich}, {Veredas}, {Wagner}, {Winkler}, {Zapatero Osorio}, {Zechmeister}, {Abell{\'a}n de Paco}, {Anglada-Escud{\'e}}, {del Burgo}, {Klutsch}, {Lizon}, {L{\'o}pez-Morales}, {Morales}, {Perryman}, {Tulloch}, \& {Xu}}]{Quirrenbach_2014}
{Quirrenbach}, A., {Amado}, P.~J., {Caballero}, J.~A., {et~al.} 2014, in Society of Photo-Optical Instrumentation Engineers (SPIE) Conference Series, Vol. 9147, Ground-based and Airborne Instrumentation for Astronomy V, ed. S.~K. {Ramsay}, I.~S. {McLean}, \& H.~{Takami}, 91471F

\bibitem[{{Rackham} {et~al.}(2017){Rackham}, {Espinoza}, {Apai}, {L{\'o}pez-Morales}, {Jord{\'a}n}, {Osip}, {Lewis}, {Rodler}, {Fraine}, {Morley}, \& {Fortney}}]{Rackham_2017}
{Rackham}, B., {Espinoza}, N., {Apai}, D., {et~al.} 2017, \apj, 834, 151

\bibitem[{{Rajpurohit} {et~al.}(2013){Rajpurohit}, {Reylé, C.}, {Allard, F.}, {Homeier, D.}, {Schultheis, M.}, {Bessell, M. S.}, \& {Robin, A. C.}}]{Rajpurohit_2013}
{Rajpurohit}, A.~S., {Reylé, C.}, {Allard, F.}, {et~al.} 2013, A\&A, 556, A15

\bibitem[{{Redfield} {et~al.}(2024){Redfield}, {Batalha}, {Benneke}, {Biller}, {Espinoza}, {France}, {Konopacky}, {Kreidberg}, {Rauscher}, \& {Sing}}]{Redfield_2024}
{Redfield}, S., {Batalha}, N., {Benneke}, B., {et~al.} 2024, arXiv e-prints, arXiv:2404.02932

\bibitem[{{Ricker} {et~al.}(2015){Ricker}, {Winn}, {Vanderspek}, {Latham}, {Bakos}, {Bean}, {Berta-Thompson}, {Brown}, {Buchhave}, {Butler}, {Butler}, {Chaplin}, {Charbonneau}, {Christensen-Dalsgaard}, {Clampin}, {Deming}, {Doty}, {De Lee}, {Dressing}, {Dunham}, {Endl}, {Fressin}, {Ge}, {Henning}, {Holman}, {Howard}, {Ida}, {Jenkins}, {Jernigan}, {Johnson}, {Kaltenegger}, {Kawai}, {Kjeldsen}, {Laughlin}, {Levine}, {Lin}, {Lissauer}, {MacQueen}, {Marcy}, {McCullough}, {Morton}, {Narita}, {Paegert}, {Palle}, {Pepe}, {Pepper}, {Quirrenbach}, {Rinehart}, {Sasselov}, {Sato}, {Seager}, {Sozzetti}, {Stassun}, {Sullivan}, {Szentgyorgyi}, {Torres}, {Udry}, \& {Villasenor}}]{Ricker_2015}
{Ricker}, G.~R., {Winn}, J.~N., {Vanderspek}, R., {et~al.} 2015, JATIS, 1, 014003

\bibitem[{Rivera {et~al.}(2005)Rivera, Lissauer, Butler, Marcy, Vogt, Fischer, Brown, Laughlin, \& Henry}]{Rivera_2005}
Rivera, E.~J., Lissauer, J.~J., Butler, R.~P., {et~al.} 2005, \apj, 634, 625

\bibitem[{{Rothman} {et~al.}(2010){Rothman}, {Gordon}, {Barber}, {Dothe}, {Gamache}, {Goldman}, {Perevalov}, {Tashkun}, \& {Tennyson}}]{Rothman_2010}
{Rothman}, L.~S., {Gordon}, I.~E., {Barber}, R.~J., {et~al.} 2010, \jqsrt, 111, 2139

\bibitem[{Salvatier {et~al.}(2016)Salvatier, Wiecki, \& Fonnesbeck}]{Salvatier_2016}
Salvatier, J., Wiecki, T.~V., \& Fonnesbeck, C. 2016, PeerJ Computer Science, 2, e55

\bibitem[{Schlawin {et~al.}(2020)Schlawin, Leisenring, Misselt, Greene, McElwain, Beatty, \& Rieke}]{Schlawin_2020}
Schlawin, E., Leisenring, J., Misselt, K., {et~al.} 2020, \aj, 160, 231

\bibitem[{Schwarz(1978)}]{Schwarz_1978}
Schwarz, G. 1978, The Annals of Statistics, 6, 461

\bibitem[{{Segura} {et~al.}(2010){Segura}, {Walkowicz}, {Meadows}, {Kasting}, \& {Hawley}}]{Segura_2010}
{Segura}, A., {Walkowicz}, L.~M., {Meadows}, V., {Kasting}, J., \& {Hawley}, S. 2010, Astrobiology, 10, 751

\bibitem[{{Seifahrt} {et~al.}(2018){Seifahrt}, {St{\"u}rmer}, {Bean}, \& {Schwab}}]{Seifahrt_2018}
{Seifahrt}, A., {St{\"u}rmer}, J., {Bean}, J.~L., \& {Schwab}, C. 2018, in Society of Photo-Optical Instrumentation Engineers (SPIE) Conference Series, Vol. 10702, Ground-based and Airborne Instrumentation for Astronomy VII, ed. C.~J. {Evans}, L.~{Simard}, \& H.~{Takami}, 107026D

\bibitem[{{Shields} {et~al.}(2016){Shields}, {Ballard}, \& {Johnson}}]{Shields_2016}
{Shields}, A.~L., {Ballard}, S., \& {Johnson}, J.~A. 2016, \physrep, 663, 1

\bibitem[{{Smithsonian Astrophysical Observatory}(2000)}]{sao_2000}
{Smithsonian Astrophysical Observatory}. 2000, {SAOImage DS9: A utility for displaying astronomical images in the X11 window environment}, Astrophysics Source Code Library, record ascl:0003.002

\bibitem[{{Sneep} \& {Ubachs}(2005)}]{Sneep_2005}
{Sneep}, M. \& {Ubachs}, W. 2005, \jqsrt, 92, 293

\bibitem[{Tamburo {et~al.}(2022)Tamburo, Muirhead, McCarthy, Hart, Gracia, Vos, Bardalez~Gagliuffi, Faherty, Theissen, Agol, Skinner, \& Sagear}]{Tamburo_2022}
Tamburo, P., Muirhead, P.~S., McCarthy, A.~M., {et~al.} 2022, \aj, 163, 253

\bibitem[{{Tamburo} {et~al.}(2022){Tamburo}, {Muirhead}, {McCarthy}, {Hart}, {Vos}, {Agol}, {Theissen}, {Gracia}, {Bardalez Gagliuffi}, \& {Faherty}}]{Tamburo_2022b}
{Tamburo}, P., {Muirhead}, P.~S., {McCarthy}, A.~M., {et~al.} 2022, \aj, 164, 252

\bibitem[{{Tarter} {et~al.}(2007){Tarter}, {Backus}, {Mancinelli}, {Aurnou}, {Backman}, {Basri}, {Boss}, {Clarke}, {Deming}, {Doyle}, {Feigelson}, {Freund}, {Grinspoon}, {Haberle}, {Hauck}, {Heath}, {Henry}, {Hollingsworth}, {Joshi}, {Kilston}, {Liu}, {Meikle}, {Reid}, {Rothschild}, {Scalo}, {Segura}, {Tang}, {Tiedje}, {Turnbull}, {Walkowicz}, {Weber}, \& {Young}}]{Tarter_2007}
{Tarter}, J.~C., {Backus}, P.~R., {Mancinelli}, R.~L., {et~al.} 2007, Astrobiology, 7, 30

\bibitem[{{Thalman} {et~al.}(2014){Thalman}, {Zarzana}, {Tolbert}, \& {Volkamer}}]{Thalman_2014}
{Thalman}, R., {Zarzana}, K.~J., {Tolbert}, M.~A., \& {Volkamer}, R. 2014, \jqsrt, 147, 171

\bibitem[{{Theano Development Team}(2016)}]{exoplanet:theano}
{Theano Development Team}. 2016, arXiv e-prints, abs/1605.02688

\bibitem[{{Tian}(2009)}]{Tian_2009}
{Tian}, F. 2009, \apj, 703, 905

\bibitem[{{Tian} \& {Heng}(2024)}]{Tian_2023}
{Tian}, M. \& {Heng}, K. 2024, \apj, 963, 157

\bibitem[{{Valencia} {et~al.}(2007){Valencia}, {Sasselov}, \& {O'Connell}}]{Valencia_2007}
{Valencia}, D., {Sasselov}, D.~D., \& {O'Connell}, R.~J. 2007, \apj, 656, 545

\bibitem[{Vehtari {et~al.}(2016)Vehtari, Gelman, \& Gabry}]{Vehtari_2016}
Vehtari, A., Gelman, A., \& Gabry, J. 2016, Statistics and Computing, 27, 1413

\bibitem[{{Vidotto} {et~al.}(2014){Vidotto}, {Gregory}, {Jardine}, {Donati}, {Petit}, {Morin}, {Folsom}, {Bouvier}, {Cameron}, {Hussain}, {Marsden}, {Waite}, {Fares}, {Jeffers}, \& {do Nascimento}}]{Vidotto_2014}
{Vidotto}, A.~A., {Gregory}, S.~G., {Jardine}, M., {et~al.} 2014, \mnras, 441, 2361

\bibitem[{{Vidotto} {et~al.}(2013){Vidotto}, {Jardine}, {Morin}, {Donati}, {Lang}, \& {Russell}}]{Vidotto_2013}
{Vidotto}, A.~A., {Jardine}, M., {Morin}, J., {et~al.} 2013, \aap, 557, A67

\bibitem[{Virtanen {et~al.}(2020)Virtanen, Gommers, Oliphant, Haberland, Reddy, Cournapeau, Burovski, Peterson, Weckesser, Bright, {van der Walt}, Brett, Wilson, Millman, Mayorov, Nelson, Jones, Kern, Larson, Carey, Polat, Feng, Moore, {VanderPlas}, Laxalde, Perktold, Cimrman, Henriksen, Quintero, Harris, Archibald, Ribeiro, Pedregosa, {van Mulbregt}, \& {SciPy 1.0 Contributors}}]{scipy_2020}
Virtanen, P., Gommers, R., Oliphant, T.~E., {et~al.} 2020, Nature Methods, 17, 261

\bibitem[{{Wachiraphan} {et~al.}(2024){Wachiraphan}, {Berta-Thompson}, {Diamond-Lowe}, {Winters}, {Murray}, {Zhang}, {Xue}, {Morley}, {Rosario-Franco}, \& {Duvvuri}}]{Wachiraphan_2024}
{Wachiraphan}, P., {Berta-Thompson}, Z.~K., {Diamond-Lowe}, H., {et~al.} 2024, \aj, arXiv:2410.10987, submitted

\bibitem[{Waskom(2021)}]{Waskom_2021}
Waskom, M.~L. 2021, JOSS, 6, 3021

\bibitem[{{Weiner Mansfield} {et~al.}(2024){Weiner Mansfield}, {Xue}, {Zhang}, {Mahajan}, {Ih}, {Koll}, {Bean}, {Coy}, {Eastman}, {Kempton}, \& {Kite}}]{Mansfield_2024}
{Weiner Mansfield}, M., {Xue}, Q., {Zhang}, M., {et~al.} 2024, \apjl, 975, L22

\bibitem[{Winn {et~al.}(2011)Winn, Matthews, Dawson, Fabrycky, Holman, Kallinger, Kuschnig, Sasselov, Dragomir, Guenther, Moffat, Rowe, Rucinski, \& Weiss}]{Winn_2011}
Winn, J.~N., Matthews, J.~M., Dawson, R.~I., {et~al.} 2011, \apj, 737, L18

\bibitem[{{Wordsworth}(2015)}]{Wordsworth_2015}
{Wordsworth}, R. 2015, \apj, 806, 180

\bibitem[{{Xue} {et~al.}(2024){Xue}, {Bean}, {Zhang}, {Mahajan}, {Ih}, {Eastman}, {Lunine}, {Mansfield}, {Coy}, {Kempton}, {Koll}, \& {Kite}}]{Xue_2024}
{Xue}, Q., {Bean}, J.~L., {Zhang}, M., {et~al.} 2024, \apjl, 973, L8

\bibitem[{Yao {et~al.}(2018)Yao, Vehtari, Simpson, \& Gelman}]{Yao_2018}
Yao, Y., Vehtari, A., Simpson, D., \& Gelman, A. 2018, Bayesian Analysis, 13, 917

\bibitem[{Zhang {et~al.}(2024)Zhang, Hu, Inglis, Dai, Bean, Knutson, Lam, Goffo, \& Gandolfi}]{Zhang_2024}
Zhang, M., Hu, R., Inglis, J., {et~al.} 2024, \apjl, 961, L44

\bibitem[{{Zieba} {et~al.}(2023){Zieba}, {Kreidberg}, {Ducrot}, {Gillon}, {Morley}, {Schaefer}, {Tamburo}, {Koll}, {Lyu}, {Acu{\~n}a}, {Agol}, {Iyer}, {Hu}, {Lincowski}, {Meadows}, {Selsis}, {Bolmont}, {Mandell}, \& {Suissa}}]{Zieba_2023}
{Zieba}, S., {Kreidberg}, L., {Ducrot}, E., {et~al.} 2023, \nat, 620, 746

\bibitem[{{Zilinskas} {et~al.}(2022){Zilinskas}, {van Buchem}, {Miguel}, {Louca}, {Lupu}, {Zieba}, \& {van Westrenen}}]{Zilinskas_2022}
{Zilinskas}, M., {van Buchem}, C.~P.~A., {Miguel}, Y., {et~al.} 2022, \aap, 661, A126

\end{thebibliography}

\begin{appendix}
\label{section_appendix}

\section{Raw photometry}
\label{appendix:raw photometry}

We present the absolute measurements for each visit before normalising. The average absolute flux is  in visit 1, 2 and 3, respectively. Overall, the values are as expected for the target, as reported in Sect. \ref{subsection:individual fit}. 

\begin{figure}[ht]
\centering
\subfloat{\includegraphics[width = 0.42\textwidth]{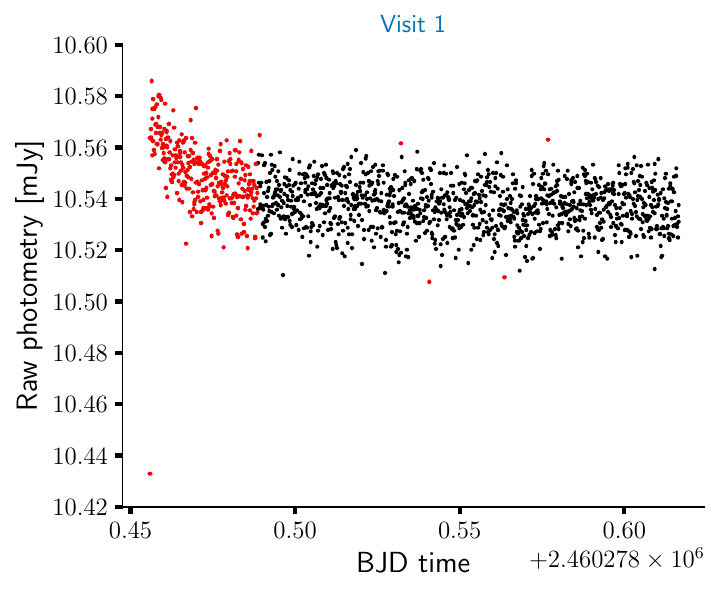}}
\newline
\subfloat{\includegraphics[width = 0.42\textwidth]{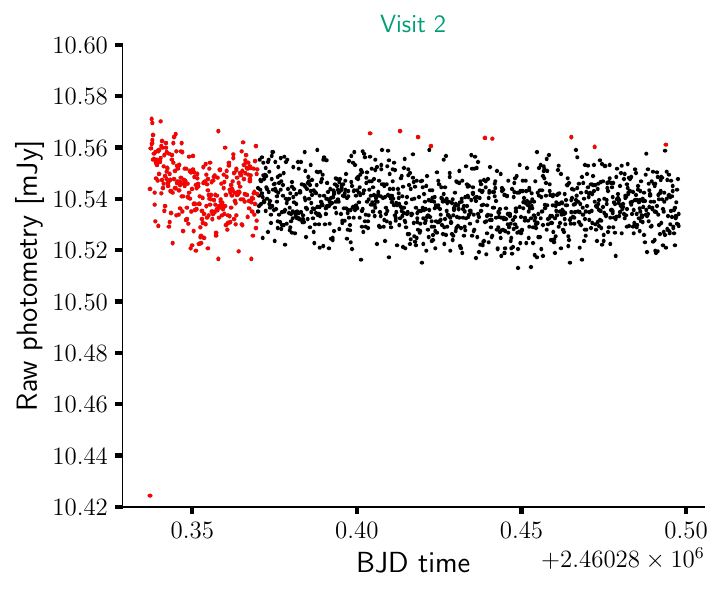}}
\newline
\raggedright
\subfloat{\includegraphics[width = 0.42\textwidth]{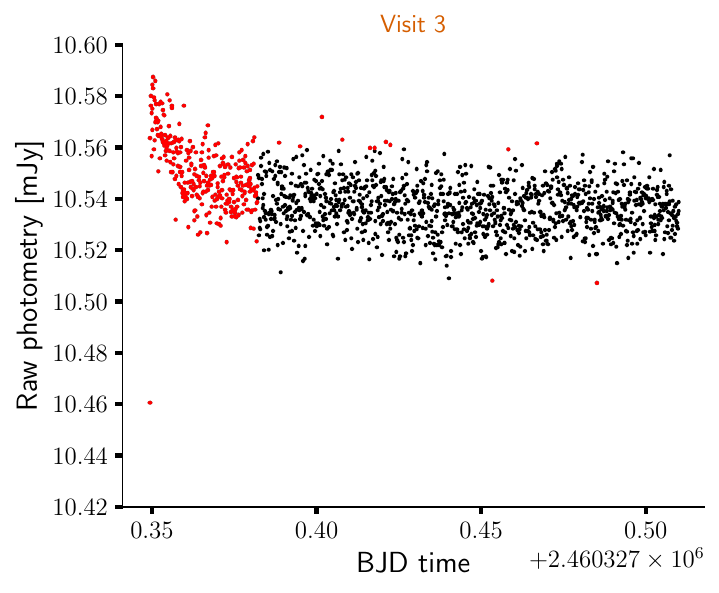}}

\caption{Raw photometry for visit 1 (top), visit 2 (middle) and visit 3 (bottom) in units of mJy as a function of time. Red data points are discarded data points as described in Sect. \ref{subsection:analysis}.}
\label{fig:raw photometry}
\end{figure}
\FloatBarrier

\section{Allan variance plots}
\label{app: rms}

\begin{figure}
\centering
\subfloat[Visit 1]{\includegraphics[width = 0.35\textwidth]{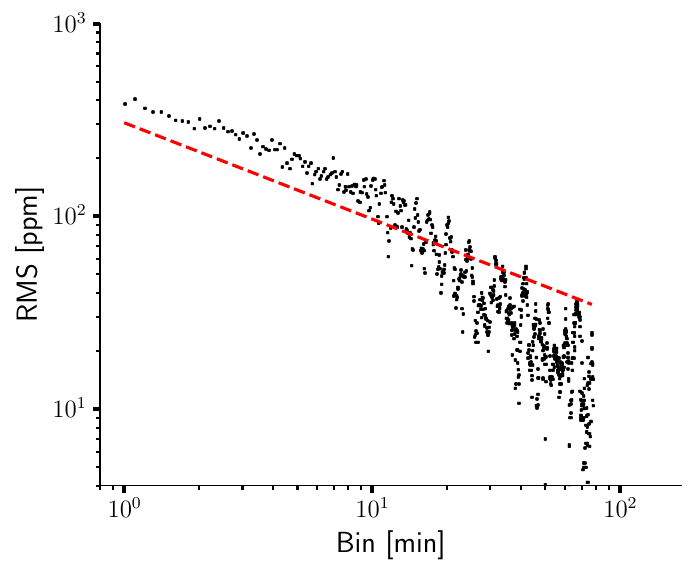}}
\newline
\subfloat[Visit 2]{\includegraphics[width = 0.35\textwidth]{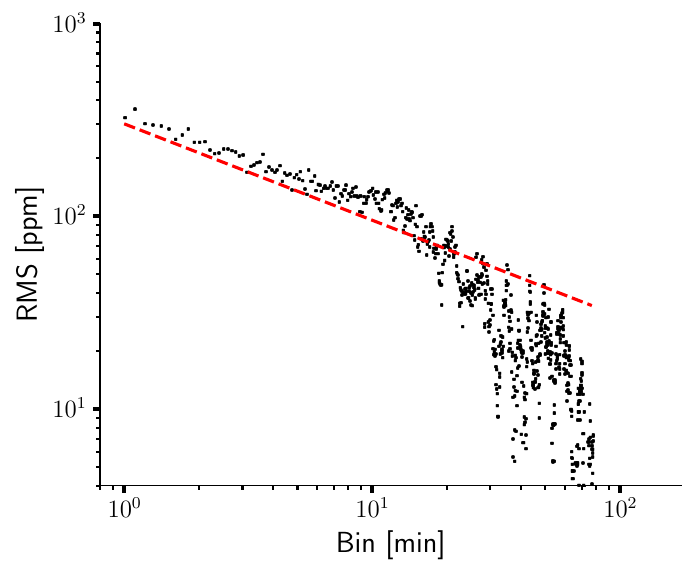}}
\newline
\raggedright
\subfloat[Visit 3]{\includegraphics[width = 0.35\textwidth]{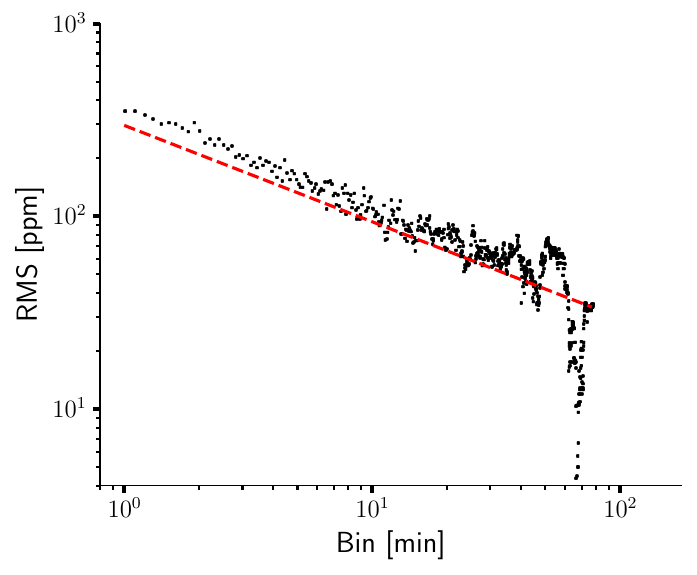}}

\caption{Photometric residual RMS as a function of bin size in minutes. The red dashed line shows the expected Poissonian noise precision normalised to the unbinned RMS. The panel on the left corresponds to visit 1, the middle panel to visit 2 and the right panel to visit 3.}
\label{fig:rms}
\end{figure}

Here we present the root mean square (RMS) of the residual measurements as a function of bin size. In the absence of correlated noise, the residual RMS decreases as 1/$\sqrt{n}$, where $n$ is the size of the bin. The resulting Allan variance plots \citep{Allan_1966} of the residuals for each visit are shown in Fig. \ref{fig:rms}. As mentioned in Sec. \ref{subsection:analysis}, the detrending models for all visits consist of a linear trend in time and background. 

\FloatBarrier

\section{Atmospheric models for heat redistribution boundary scenarios}
\label{app: atmospheres}

We present the atmospheric models for the cases of no heat redistribution to the nightside (f=2/3) and full heat redistribution (f=1/4) described in Sect. \ref{subsection:atmospheric models} for a pure CO$_{2}$ atmosphere, pure H$_{2}$O, 10\% CO$_{2}$ and 90\% H$_{2}$O and 90\% CO$_{2}$ and 10\% H$_{2}$O atmosphere. In all cases the surface albedo and pressure are 0.1 and 1000 mbar, respectively. 

\begin{figure}
    \centering
    \resizebox{0.99\hsize}{!}{\includegraphics{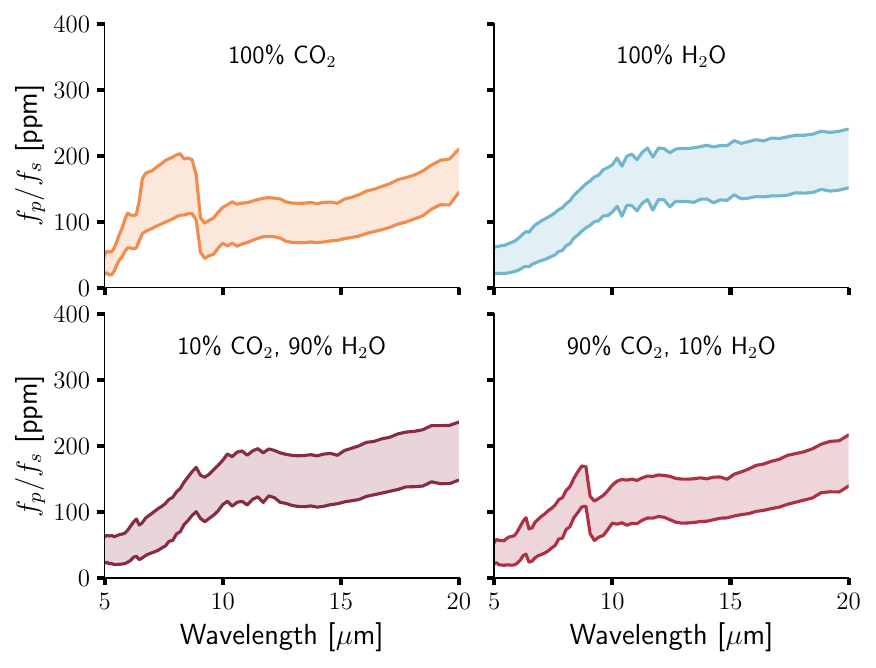}}
    \caption{Emission spectra for different atmospheric scenarios. The curves show the planetary flux to stellar flux ratio in ppm as a function of wavelength in $\mu$m. Each panel corresponds to the atmospheric composition labelled on the top of the panel. In each panel, the top curve represents the model for no heat redistribution (f=2/3), while the bottom curve is for full redistribution (f=1/4). The shaded area covers the scenarios in between the boundary cases.}
    \label{fig:extreme f cases}
\end{figure}

\section{Analysis for different aperture sizes}
\label{app: Aperture sizes}

Here we present the analysis following Sect. \ref{section:method} for apertures between 3 and 20 pixels for photometric extraction. In each case, we compute the MAD of the raw light-curve and for the trimmed data after removing the first segment, the posterior estimate on the occultation depth and the residual RMS. The resulting values are presented in Table \ref{tab:apertures}. The selection of aperture size for the main analysis is based on these values. In all cases, the background annulus was fixed with an inner radius of 30 pixels and outer radius of 50 pixels.

\begin{table*}[!hbt]
\centering\setstretch{1.0}
        \caption{Analysis for different aperture sizes.}
        \begin{tabular}{c || c | c | c | c | c | c}
        \hline
        \hline
            Aperture size & MAD raw data & MAD trimmed data & $f_{p}/f_{s, \ joint}$ & $f_{p}/f_{s, \ i}$ & RRMS, joint & RRMS, i \\
            \hline
             & 1211 & 959 & & 321 $\pm$ 56 & & 849  \\
            3 & 1231 & 923 & 410 $\pm$ 32 & 374 $\pm$ 57  & 833 & 843 \\
             & 1210 & 878 & & 533 $\pm$ 57 & & 804 \\
            \hline
             & 2033 & 853 & & 271 $\pm$ 62 & & 797 \\
            4 & 943 & 807 & 341 $\pm$ 34 & 336 $\pm$ 53 & 770 & 762 \\
             & 1158 & 850 & & 401 $\pm$ 54 & & 757 \\
            \hline
             & 924 & 841 & & 239 $\pm$ 52 & & 778 \\
            \textbf{5} & 885 & 799 & 311 $\pm$ 31 & 337 $\pm$ 53 & 767 & 768 \\
             & 985 & 802 & & 357 $\pm$ 52 & & 754 \\
            \hline
             & 949 & 841 & & 210 $\pm$ 54 & & 795 \\
            6 & 904 & 832 & 279 $\pm$ 32 & 312 $\pm$ 54 & 799 & 805 \\
             & 960 & 831 & & 321 $\pm$ 54 & & 796 \\
            \hline
             & 979 & 851 & & 215 $\pm$ 62 & & 817 \\
            7 & 1310 & 1306 & 225 $\pm$ 37 & 131 $\pm$ 66 & 960 & 1203 \\
             & 1011 & 833 & & 333 $\pm$ 64 & & 802 \\
            \hline
             & 947 & 862 & & 231 $\pm$ 58 & & 822 \\
            8 & 963 & 919 & 273 $\pm$ 33 & 293 $\pm$ 57 & 846 & 863 \\
             & 1064 & 882 & & 295 $\pm$ 57 & & 852 \\
            \hline
             & 949 & 840 & & 198 $\pm$ 64 & & 818 \\
            9 & 968 & 894 & 283 $\pm$ 37 & 295 $\pm$ 61 & 920 & 861 \\
             & 1344 & 1197 & & 356 $\pm$ 65 & & 1060 \\
            \hline
             & 964 & 869 & & 203 $\pm$ 61 & & 851 \\
            10 & 1003 & 928 & 273 $\pm$ 35 & 311 $\pm$ 61 & 893 & 892 \\
             & 1142 & 998 & & 305 $\pm$ 64 & & 934 \\
            \hline
             & 1019 & 927 & & 179 $\pm$ 63 & & 890 \\
            11 & 1074 & 976 & 272 $\pm$ 37 & 305 $\pm$ 63 & 932 & 930 \\
             & 1171 & 1036 & & 332 $\pm$ 61 & & 972 \\
            \hline
             & 1090 & 992 & & 187 $\pm$ 65 & & 953 \\
            12 & 1104 & 1059 & 287 $\pm$ 38 & 336 $\pm$ 68 & 985 & 977 \\
             & 1242 & 1093 & & 339 $\pm$ 68 & & 1024 \\
            \hline
             & 1130 & 1065 & & 185 $\pm$ 69 & & 1003 \\
            13 & 1196 & 1143 & 285 $\pm$ 42 & 346 $\pm$ 70 & 1032 & 1036 \\
             & 1246 & 1118 & & 327 $\pm$ 71 & & 1055 \\
            \hline
             & 1196 & 1132 & & 190 $\pm$ 74 & & 1051 \\
            14 & 1230 & 1195 & 259 $\pm$ 42 & 338 $\pm$ 72 & 1084 & 1082 \\
             & 1310 & 1186 & & 250 $\pm$ 77 & & 1117 \\
            \hline
             & 1264 & 1156 & & 177 $\pm$ 76 & & 1089 \\
            15 & 1302 & 1224 & 243 $\pm$ 45 & 300 $\pm$ 76 & 1121 & 1109 \\
             & 1390 & 1233 & & 257 $\pm$ 77 & & 1163 \\
            \hline
             & 1314 & 1272 & & 191 $\pm$ 82 & & 1160 \\
            16 & 1375 & 1312 & 235 $\pm$ 47 & 283 $\pm$ 81 & 1183 & 1168 \\
             & 1427 & 1311 & & 232 $\pm$ 81 & & 1221\\
            \hline
             & 1346 & 1244 & & 190 $\pm$ 84 & & 1200 \\
            17 & 1423 & 1316 & 201 $\pm$ 48 & 245 $\pm$ 83 & 1226 & 1227 \\
             & 1508 & 1357 & & 167 $\pm$ 85 & & 1250 \\
            \hline
             & 1464 & 1402 & & 194 $\pm$ 87 & & 1272 \\
            18 & 1463 & 1423 & 213 $\pm$ 51 & 269 $\pm$ 88 & 1285 & 1274 \\
             & 1543 & 1410 & & 174 $\pm$ 85 & & 1310 \\
            \hline
             & 1537 & 1446 & & 196 $\pm$ 92 & & 1324 \\
            19 & 1535 & 1475 & 221 $\pm$ 53 & 265 $\pm$ 94 & 1344 & 1325 \\
             & 1614 & 1416 & & 201 $\pm$ 91 & & 1385 \\
            \hline
             & 1627 & 1537 & & 162 $\pm$ 94 & & 1382 \\
            20 & 1603 & 1520 & 197 $\pm$ 56 & 252 $\pm$ 94 & 1401 & 1381 \\
             & 1734 & 1543 & & 180 $\pm$ 94 & & 1442 \\
            \hline 
   
            \hline
        \end{tabular}
        \label{tab:apertures}
        \tablefoot{Median absolute deviation (MAD) of the raw data for each visit, MAD of the trimmed datasets after removing the first 47 minutes, posterior estimate on the occultation depth and residual RMS for different aperture sizes. The smallest circular aperture has a radius of 3 pixels, in increment of one pixel up to the largest radius of 20 pixels. Bold text on the leftmost column highlights the aperture size selected for the main analysis. All values have units of ppm. Cells with multiple entries correspond to the visits ordered from top to bottom.}
\end{table*}

\section{Corner plot of the joint analysis}
\label{appendix:corner plots}

\begin{figure*}
    \centering
    \resizebox{\hsize}{!}{\includegraphics{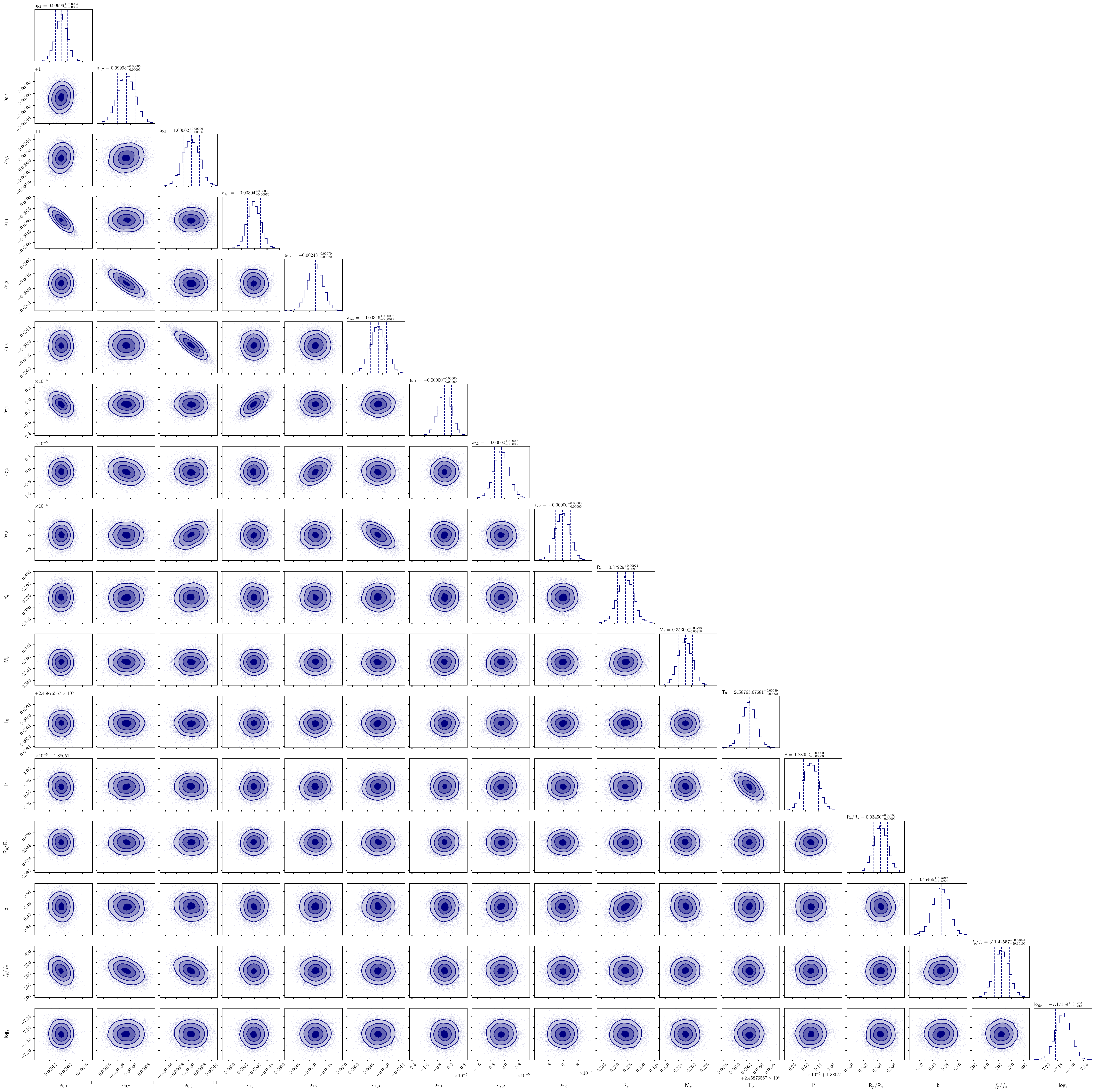}}
    \caption{Posterior distributions and joint correlations plot corresponding to to joint analysis of TOI-1468 b  observations.}
    \label{fig:corner joint}
\end{figure*}

Figure \ref{fig:corner joint} shows the full joint posterior correlation plots of the joint analysis. $a_{0,j}$ is the mean flux, $a_{1,j}$ the slope of the linear trend in time and $a_{6,j}$ the background level of visit $j$ with $j \in \{1,2,3\}$; $R_{s}$ and $M_{s}$ are the stellar radius and mass, $T_{0}$ is the mid-transit time, $P$ the orbital period, $R_{p}/R_{s}$ is the ratio of the planetary radius to the stellar radius, $b$ is the impact parameter, $f_{p}/f_{s}$ the occultation depth, and $\log(s)$ is the natural logarithm of the flux uncertainty for each measurement.

\section{\texttt{EXOFASTv2} global fit}
\label{appendix:exofast}

Table \ref{tab:exofast} presents the best-fit parameters of the global fit performed with \texttt{EXOFASTv2} \citep{Eastman_2019} described in Sect. \ref{subsection:analysis}.


\setlength{\tabcolsep}{0.2pt}
\onecolumn
\begin{longtable}{lccccc}
\caption{\texttt{EXOFASTv2} median values and 68\% confidence interval for TOI-1468.} \\
\hline\hline
Parameter & Description & & Values & & \\
\hline
\vspace{5mm}
\endfirsthead 
\caption{continued.}\\ 
\hline\hline 
Parameter & Description & & Values & & \\
\hline  
\endhead
\hline
\endfoot
Stellar parameters: &  &  &  &  & \\
$M_*$\dotfill & Mass (\msun)\dotfill & $0.3527^{+0.0081}_{-0.0079}$ & & & \\
$R_*$\dotfill & Radius (\rsun)\dotfill & $0.3714^{+0.0100}_{-0.010}$ & & & \\
$L_*$\dotfill & Luminosity (\lsun)\dotfill & $0.01614^{+0.00054}_{-0.00052}$ & & & \\
$F_{Bol}$\dotfill & Bolometric Flux (cgs)\dotfill &$8.45^{+0.28}_{-0.27} \times 10^{-10}$ & & & \\
$\rho_*$\dotfill & Density (cgs)\dotfill &$9.70^{+0.85}_{-0.76}$ & & & \\
$\log{g}$\dotfill & Surface gravity (cgs)\dotfill &$4.846\pm0.025$ & & & \\
$T_{\rm eff}$\dotfill & Effective temperature (K)\dotfill &$3376^{+45}_{-44}$ & & & \\
$[{\rm Fe/H}]$\dotfill & Metallicity (dex)\dotfill &$-0.01^{+0.21}_{-0.22}$ & & & \\
$K_S$\dotfill & Absolute Ks-band mag (mag)\dotfill &$6.530^{+0.020}_{-0.021}$ & & & \\
$k_S$\dotfill & Apparent Ks-band mag (mag)\dotfill &$8.494\pm0.020$ & & & \\
$A_V$\dotfill & V-band extinction (mag)\dotfill &$0.051^{+0.046}_{-0.036}$ & & & \\
$\sigma_{SED}$\dotfill & SED photometry error scaling \dotfill &$2.25^{+0.85}_{-0.54}$ & & & \\
$\varpi$\dotfill & Parallax (mas)\dotfill &$40.462^{+0.076}_{-0.075}$ & & & \\
\vspace{5mm}
$d$\dotfill & Distance (pc)\dotfill &$24.714\pm0.046$ & & & \\
Planetary parameters: &  & b & c &  & \\
$P$\dotfill & Period (days)\dotfill &$1.8805201^{+0.0000025}_{-0.0000030}$&$15.532477^{+0.000025}_{-0.000026}$ & & \\
$R_P$\dotfill & Radius (\re)\dotfill &$1.401^{+0.059}_{-0.060}$&$2.126^{+0.092}_{-0.093}$ & & \\
$M_P$\dotfill & Mass (\me)\dotfill &$3.04\pm0.46$&$4.1\pm1.1$ & & \\
$T_C$\dotfill & Observed Time of conjunction\tablefootmark{(1)} (\bjdtdb)\dotfill &$2458765.67755^{+0.00097}_{-0.00081}$&$2458766.9267\pm0.0012$ & & \\
$T_0$\dotfill & Obs time of min proj sep (\bjdtdb)\dotfill &$2459264.01539\pm0.00054$&$2459419.29076^{+0.00058}_{-0.00061}$ & & \\
$a$\dotfill & Semi-major axis (AU)\dotfill &$0.02107\pm0.00016$&$0.08608\pm0.00065$ & & \\
$i$\dotfill & Inclination (Degrees)\dotfill &$87.82^{+0.42}_{-0.34}$&$89.220^{+0.14}_{-0.062}$ & & \\
$e$\dotfill & Eccentricity \dotfill &$0.0099^{+0.018}_{-0.0050}$&$0.111^{+0.15}_{-0.084}$ & & \\
$\omega_*$\dotfill & Arg of periastron (Degrees)\dotfill &$-170^{+69}_{-82}$&$130^{+100}_{-73}$ & & \\
$\dot{\omega}_{\rm GR}$\dotfill & Computed GR precession ($^\circ$/century)\dotfill &$3.468^{+0.053}_{-0.052}$&$0.1046^{+0.0059}_{-0.0024}$ & & \\
$T_{\rm eq}$\dotfill & Equilibrium temperature\tablefootmark{(2)} (K)\dotfill &$683.5^{+6.3}_{-6.2}$&$338.1\pm3.1$ & & \\
$\tau_{\rm circ}$\dotfill & Tidal circ timescale (Gyr)\dotfill &$7.3^{+2.2}_{-1.7}$&$8600^{+4300}_{-4000}$ & & \\
$K$\dotfill & RV semi-amplitude (m/s)\dotfill &$3.16^{+0.48}_{-0.47}$&$2.12^{+0.57}_{-0.58}$ & & \\
$R_P/R_*$\dotfill & Radius of planet in stellar radii \dotfill &$0.03461^{+0.00092}_{-0.00100}$&$0.0526^{+0.0015}_{-0.0017}$ & & \\
$a/R_*$\dotfill & Semi-major axis in stellar radii \dotfill &$12.20^{+0.35}_{-0.33}$&$49.8^{+1.4}_{-1.3}$ & & \\
$\delta$\dotfill & $\left(R_P/R_*\right)^2$ \dotfill &$0.001198^{+0.000065}_{-0.000068}$&$0.00276^{+0.00016}_{-0.00018}$ & & \\
$\delta_{\rm TESS}$\dotfill & Transit depth in TESS (frac)\dotfill &$0.00133^{+0.00018}_{-0.00011}$&$0.00291^{+0.00025}_{-0.00019}$ & & \\
$\tau$\dotfill & In/egress transit duration (days)\dotfill &$0.00193^{+0.00016}_{-0.00015}$&$0.0067^{+0.0015}_{-0.0022}$ & & \\
$T_{14}$\dotfill & Total transit duration (days)\dotfill &$0.0454^{+0.0016}_{-0.0011}$&$0.0775^{+0.0020}_{-0.0021}$ & & \\
$T_{FWHM}$\dotfill & FWHM transit duration (days)\dotfill &$0.0435^{+0.0017}_{-0.0012}$&$0.0708^{+0.0018}_{-0.0017}$ & & \\
$b$\dotfill & Transit impact parameter \dotfill &$0.465^{+0.061}_{-0.081}$&$0.664^{+0.069}_{-0.22}$ & & \\
$b_S$\dotfill & occultation impact parameter \dotfill &$0.464^{+0.061}_{-0.082}$&$0.686^{+0.057}_{-0.094}$ & & \\
$\tau_S$\dotfill & In/egress occultation duration (days)\dotfill &$0.00192^{+0.00017}_{-0.00016}$&$0.00743^{+0.0012}_{-0.00098}$ & & \\
$T_{S,14}$\dotfill & Total occultation duration (days)\dotfill &$0.0454^{+0.0014}_{-0.0012}$&$0.0791^{+0.017}_{-0.0027}$ & & \\
$T_{S,FWHM}$\dotfill & FWHM occultation duration (days)\dotfill &$0.0434^{+0.0015}_{-0.0012}$&$0.0715^{+0.017}_{-0.0025}$ & & \\
$\rho_P$\dotfill & Density (cgs)\dotfill &$6.1^{+1.3}_{-1.1}$&$2.31^{+0.72}_{-0.66}$ & & \\
$logg_P$\dotfill & Surface gravity (cgs)\dotfill &$3.181^{+0.074}_{-0.080}$&$2.94^{+0.11}_{-0.14}$ & & \\
$\Theta$\dotfill & Safronov Number \dotfill &$0.0091^{+0.0015}_{-0.0014}$&$0.0329^{+0.0089}_{-0.0090}$ & & \\
$\fave$\dotfill & Incident Flux (\fluxcgs)\dotfill &$0.0495\pm0.0018$&$0.00290^{+0.00013}_{-0.00016}$ & & \\
$T_S$\dotfill & Observed Time of occultation (\bjdtdb)\dotfill &$2458766.6124^{+0.0026}_{-0.0022}$&$2458774.52^{+0.53}_{-1.3}$ & & \\
$e\cos{\omega_*}$\dotfill & \dotfill & $-0.0044^{+0.0017}_{-0.0013}$&$-0.018^{+0.053}_{-0.13}$ & & \\
$e\sin{\omega_*}$\dotfill & \dotfill & $-0.001^{+0.014}_{-0.017}$&$0.032^{+0.17}_{-0.073}$ & & \\
$M_P\sin i$\dotfill & Minimum mass (\me)\dotfill &$3.04\pm0.46$&$4.1\pm1.1$ & & \\
$M_P/M_*$\dotfill & Mass ratio \dotfill &$2.59\pm0.39 \times 10^{-5}$&$3.47^{+0.93}_{-0.95} \times 10^{-5}$ & & \\
$d/R_*$\dotfill & Separation at mid transit \dotfill &$12.22^{+0.42}_{-0.40}$&$47.7^{+4.4}_{-9.0}$ & & \\
$P_T$\dotfill & A priori non-grazing transit prob \dotfill &$0.0790^{+0.0027}_{-0.0026}$&$0.0199^{+0.0046}_{-0.0017}$ & & \\
$P_{T,G}$\dotfill & A priori transit prob \dotfill &$0.0847^{+0.0029}_{-0.0028}$&$0.0221^{+0.0051}_{-0.0018}$ & & \\
$P_S$\dotfill & A priori non-grazing occultation prob \dotfill &$0.0792^{+0.0027}_{-0.0026}$&$0.0185^{+0.0016}_{-0.0022}$ & & \\
\vspace{5mm}
$P_{S,G}$\dotfill & A priori occultation prob \dotfill &$0.0849^{+0.0029}_{-0.0028}$&$0.0205^{+0.0018}_{-0.0025}$ & & \\
Wavelength parameters: &  & JWST & TESS &  & \\
$u_{1}$\dotfill & Linear limb-darkening coeff \dotfill & - & $0.24^{+0.26}_{-0.17}$ & & \\
$u_{2}$\dotfill & Quadratic limb-darkening coeff \dotfill & - &$0.23^{+0.32}_{-0.31}$ & & \\
$A_T$\dotfill & Thermal emission from the planet (ppm)\dotfill & $298\pm28$ & - & & \\
\vspace{5mm}
$\delta_{S}$\dotfill & Measured occultation depth (ppm)\dotfill & $298\pm28$ & - & & \\
Telescope parameters: & & CARMENES & MAROONXblue & \hspace{-4mm} MAROONXred & \\
$\gamma_{\rm rel}$\dotfill & Relative RV Offset (m/s)\dotfill &$-0.32\pm0.45$&$0.19^{+0.88}_{-0.86}$&$-0.40^{+0.68}_{-0.69}$ & \\
$\sigma_J$\dotfill & RV Jitter (m/s)\dotfill &$2.92^{+0.44}_{-0.39}$&$2.92^{+0.99}_{-0.77}$&$2.50^{+0.74}_{-0.54}$ & \\
\vspace{7mm}
$\sigma_J^2$\dotfill & RV Jitter Variance \dotfill &$8.5^{+2.7}_{-2.1}$&$8.5^{+6.8}_{-3.9}$&$6.2^{+4.2}_{-2.4}$ & \\
Transit parameters: & & JWST & TESS Sector 17 &  &  \\
$\sigma^{2}$\dotfill & Added Variance \dotfill &$1.47^{+0.15}_{-0.14} \times 10^{-7}$&$-7.0\pm4.4 \times 10^{-8}$& & \\
\vspace{7mm}
$F_0$\dotfill & Baseline flux \dotfill &$0.999706\pm0.000049$&$1.000026\pm0.000017$& & \\
Transit parameters (continued): & & TESS Sector 42 and 43 & TESS Sector 57 &  &  \\
$\sigma^{2}$\dotfill & Added Variance \dotfill & $-8.6\pm2.8 \times 10^{-8}$&$-8.4\pm3.4 \times 10^{-8}$& & \\
$F_0$\dotfill & Baseline flux \dotfill &$1.000001\pm0.000011$&$1.000001^{+0.000014}_{-0.000013}$& & \\
\label{tab:exofast}

\end{longtable}
\tablefoot{See Table 3 in \citet{Eastman_2019} for a detailed description of all parameters.
\tablefoottext{1}{Time of conjunction is commonly reported as the ``mid-transit time''.}
\tablefoottext{2}{Assumes zero Bond albedo and perfect redistribution.}}
\twocolumn

\end{appendix}

\end{document}